\documentclass[11pt]{amsart}

\usepackage[nocompress]{cite} 

\usepackage[normalem]{ulem}
\usepackage{subfigure}
\usepackage{caption}
\usepackage{amsaddr}
\usepackage{color}
\usepackage{enumerate}
\usepackage{bbm}
\usepackage[mathscr]{eucal}
\usepackage{verbatim}
\usepackage{hyperref}
\usepackage{graphicx}
\usepackage{mathtools}
\usepackage[margin=1in]{geometry}
\usepackage{amsmath}
\usepackage{amssymb}
\newcommand*{\QEDB}{\hfill\ensuremath{\square}}%

\newcommand{\sign}{{\rm sign}}
\newtheorem{definition}{Definition}[section]

\newtheorem{theorem}{Theorem}[section]
\newtheorem{lemma}[theorem]{Lemma}

\newtheorem{corollary}[theorem]{Corollary}

\theoremstyle{remark}
\newtheorem*{remark}{Remark}

\numberwithin{equation}{section}

\def\presuper#1#2%
  {\mathop{}%
   \mathopen{\vphantom{#2}}^{#1}%
   \kern-\scriptspace%
   #2}

\DeclareMathOperator*{\IGP}{\presuper{IGP}{\mathit{A}}}
\DeclareMathOperator*{\Tri}{\presuper{Tri}{\mathit{A}}}
\DeclareMathOperator*{\MRS}{{\rm MRS}}
\DeclareMathOperator*{\SSS}{{\rm SS}}
\newcommand{\NS}{NS}

\begin{document}

\title[Exact probabilities for press perturbations]{Exact probabilities for the indeterminacy of complex networks as perceived through press perturbations} 

\author{David Koslicki${}^{1*}$, Mark Novak${}^2$}
\address{${}^1$ Mathematics Department, Oregon State University, Corvallis, OR.}
\address{${}^2$ Department of Integrative Biology, Oregon State University, Corvallis, OR.}
\thanks{${}^*$ Corresponding Author: \url{david.koslicki@math.oregonstate.edu}}

\date{\today}
\begin{abstract}
We consider the goal of predicting how complex networks respond to chronic (press) perturbations when characterizations of their network topology and interaction strengths are associated with uncertainty. Our primary result is the derivation of exact formulas for the expected number and probability of qualitatively incorrect predictions about a system's responses under uncertainties drawn form arbitrary distributions of error. These formulas obviate the current use of simulations, algorithms, and qualitative modeling techniques. Additional indices provide new tools for identifying which links in a network are most qualitatively and quantitatively sensitive to error, and for determining the volume of errors within which predictions will remain qualitatively determinate (i.e. sign insensitive). Together with recent advances in the empirical characterization of uncertainty in ecological networks, these tools bridge a way towards probabilistic predictions of network dynamics. 
\end{abstract}
\maketitle

\section{Introduction}

The need to understand and predict how complex networks respond to perturbations of their constituent entities pervades many disciplines, including applications in communications, human health, and fisheries management \cite{Strogatz:2001fk,Ives:2007vn}.  Many of these perturbations involve sustained, chronic changes imposed on particular nodes of the network (a.k.a. `press perturbations' \cite{Bender:1984kx}), which can propagate rapidly via both direct and indirect pathways.  Ecosystem-based fisheries management, for example, needs to consider not only how alternative harvesting scenarios will alter the abundance of a particular focal species or stock, but also how such perturbations ripple through the ecosystem to affect non-targeted species \cite{Travis:2014uq}.

Predicting how networks respond to press perturbations is hindered by a number of compounding sources of uncertainty \cite{Petchey:2015fk}.  It has long been appreciated, for example, that predictions can depend crucially on knowing both the network topology and the strengths of the interactions that connect each pair of species \cite{Lawlor:1979fk, Yodzis:1988fk}.  Even low-complexity networks entail an inordinate number of indirect pathways \cite{Borrett:2003fk}.  When combined with uncertainty in a network's topology and interaction strengths, these indirect pathways can quickly render a targeted perturbation's net effects as indeterminate, leaving little predictive certainty in the magnitude or even the sign (increase or decrease) of each species' ultimate response \cite{Dambacher:2002ve, Novak:2011fj}.

Here, our goal is to understand the effect of uncertainty on the steady state variation of a complex network presumed to be well-described by a system of ordinary differential equations.  The variation of interest is either the quantitative or qualitative response of each system variable to a sustained perturbation of another system variable.  More specifically, given a system of interacting variables defined by $\tfrac{dN_i}{dt}=f_i(\vec{N})+u_i$ for $i=1,\dots,n$ variables, we are interested in determining the sign or magnitude of $\tfrac{\partial N_i}{\partial u_j}$.  In the context of ecological networks, $N_i$ is the abundance of species $i$, $f_i(\vec{N})$ is a function describing the interactions between species $i$ and a vector of other species, and $u_i$ is a scalar representing a constant rate of external input to (or removal of) species $i$.  The relationship between (the sign of) $\tfrac{\partial N_i}{\partial u_j}$ and the vector valued function $\vec{f}$ encapsulates the \textit{(sign) sensitivity} of a system's predicted dynamics to uncertainty in the species interactions.  Qualitative indeterminacy refers to the situation when the direction of a species' response cannot be predicted without quantitative knowledge of $\vec{f}$.  

Previous efforts to understand the sign sensitivity of ecological networks such as food webs have fallen into two primary categories.  Both are based on the characterization of species interactions by means of the so-called Community matrix \cite{Levins:1968yq}, with $A_{i,j}=\tfrac{\partial f_i(\vec{N})}{\partial N_j}$ reflecting a Jacobian of the system's $i=1,...,n$ growth rate equations \cite{Novak:2016zr}.  Assuming steady state conditions, the community matrix affords insight into a press perturbation $u_j$'s net propagation along all direct and indirect pathways by means of $-A^{-1}=\tfrac{\partial N_i}{\partial u_j}$ \cite{Yodzis:1988fk,Barabas:2014vn}, hereafter referred to as the Net Effects matrix.  Many other names have been ascribed to this matrix \cite{Novak:2016zr}.  Each $(i,j)$ entry of $-A^{-1}$ encapsulates a first-order approximation to the net change in species $i$'s steady state abundance due to a sustained increase in species $j$'s growth rate, assuming no bifurcations are incurred \cite{Novak:2016zr}. The elements of $-A^{-1}$ may also be normalized to understand how species respond to perturbations of their abundances, i.e. $\tfrac{\partial N_i}{\partial N_j}$ \cite{Novak:2016zr}.

Beginning with \cite{Levins:1974rt}, the first approach, typically referred to as Loop Analysis or Qualitative Modeling, has been to focus on the influence of network topology alone by specifying the elements of $A$ by their sign (i.e. $A_{i,j} = 1, 0, \text{or} -1$).  On this basis, it was reasoned in \cite{Dambacher:2002ve} that the relative frequency of positive and negative feedback loops between species provides insight into the likelihood of observing a net increase or decrease for a given species in response to a press perturbation elsewhere in its network.  That is, net effects emanating from a near equal summation of positive and negative feedback loops are inferred to be more qualitatively indeterminate than are net effects that are dominated by one or the other.  The approach is widely used \cite{Carey:2014qf,Marzloff:2016ve}, with benefits including the ability to more easily analyze alternative network topologies, and that empirical estimates of interaction strength are unnecessary.  Drawbacks include the need to compute the matrix permanent of $A$ for the total summation of feedback loops, which quickly becomes computationally challenging for large networks \cite{Jerrum:2004fr, Novak:2011fj}.  Furthermore, simulations have shown that the `weighted feedback matrix', which encapsulate the metric of positive versus negative feedback loops, quickly loses utility as its entries rapidly diminish to values of zero, implying complete indeterminacy, as network complexity increases \cite{Novak:2011fj}.  Extensions of the approach help to reduce this indeterminacy in applications where prior information or knowledge of a subset of net effects is available \cite{Hosack:2008fk, Raymond:2011bh}.

Beginning with \cite{Yodzis:1988fk}, the second approach has been to assume that network topology and some aspect of the quantitative elements of $A$ are known, and to use simulations to assess the sign sensitivity of $-A^{-1}$ to uncertainty in $A$.  For example,  \cite{Dambacher:2003ys, Melbourne-Thomas:2012uq, Raymond:2011bh} used simulations in which the values of $A$ were drawn from predefined (typically uniform) distributions, interpreting the most frequently observed sign of each entry in the resultant $-A^{-1}$ matrices as the most probable perturbation response.  Others have used simulations to compare the sign structure of $-A^{-1}$ given an assumed `true' $A$ to those produced after introducing varying degrees of error to the elements of $A$, randomly drawing these errors from log-uniform distributions \cite{Novak:2011fj, Iles:2016uq}.  The benefits of such `quantitative models' include consideration of the extreme variation of the $A_{i,j}$ magnitudes that is known to occur in nature \cite{Wootton:2005rc}, and that this approach typically exhibits less qualitative indeterminacy than do corresponding qualitative models \cite{Dambacher:2002ve, Novak:2011fj}.  However, the approach provides less clarity into the contribution of topology and requires extensive simulations or permutation tests to achieve insight.  Recently, \cite{Giordano:2016fk} have developed an algorithm for determining which entries of $A$ are sign insensitive (qualitatively determinate) to quantitative uncertainty on the assumption that a BDC decomposition of $A$ is possible.

\begin{figure}[!h]%
\begin{center}
\includegraphics[width=3.5in]{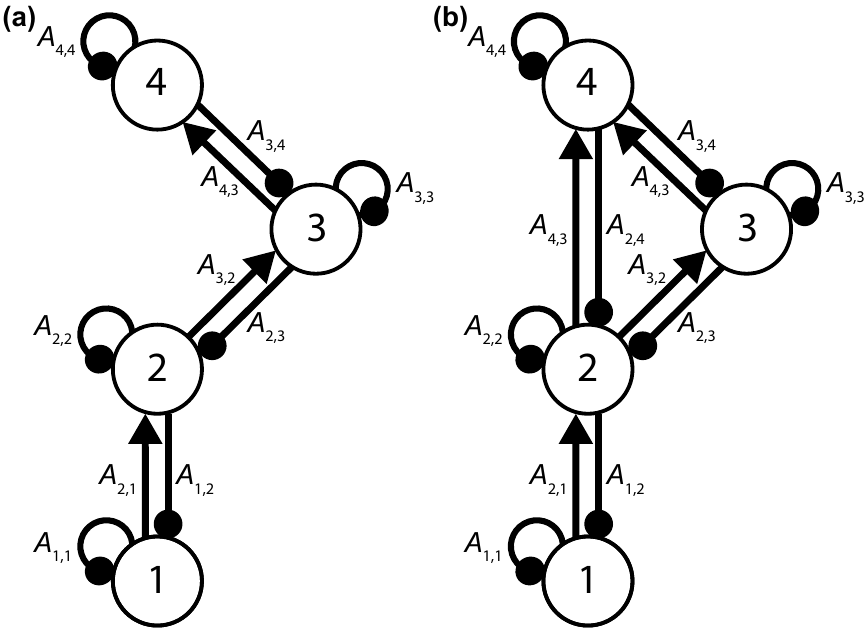}
\end{center}
\caption{(a) The four-species trophic chain (TC) network motif and (b) the intraguild predation (IGP) network motif depicted as signed digraphs. Arrowheads and circles respectively indicate the positive and negative direct effects between species as encapsulated by $A$. Following \cite{Novak:2016zr}, we let $A_{1,1}=-0.237$, $A_{2,2}=A_{3,3}=A_{4,4}=-0.015$, $A_{1,2}=A_{2,3}=A_{3,4}=-1$ and $A_{2,1}=A_{3,2}=A_{4,3}=0.1$ for both motifs, and $A_{2,4}=-1$ and $A_{4,2}=0.045$ for the IGP motif. All other entries are set to zero.}
\label{fig:motifs}%
\end{figure}

Here we provide exact formulas for the expected number and probability of making qualitatively incorrect predictions about a system's responses given by $-A^{-1}$ when the entries of $A$ are associated with error drawn from an arbitrary distribution.  The assessment of alternative network topologies, as considered by Loop Analysis, may be considered as special cases of these error distributions.  Our approach does not rest on simulations or algorithms.  Focusing on two particularly illustrative network motifs -- a four-species trophic chain (TC) and a four-species intraguild predation (IGP) motif (Fig. \ref{fig:motifs}) -- which are pervasive in food webs \cite{Stouffer:2007dn}, we provide computationally accessible methods for determining which entries of $A$ are most sensitive to error, and for determining the magnitude and volume of entry-wise errors that will incur no sign switches for any distribution of errors.  These methods enable us to demonstrate and explain the seemingly counterintuitive result that, for the parameterizations owing to \cite{Takimoto:2007ek, Novak:2016zr}, the TC motif is in fact more quantitatively sensitive than is the IGP motif, despite the TC motif being entirely sign insensitive to any error in its $A_{i,j}$ entries. We relate this result to the variance of the entries of $-A^{-1}$ as well as to the singular values and the variance of the entries of $A$, which reflect a system's asymptotic stability \cite{Allesina:2012uq}. Note that, for consistency with the mathematical literature, we henceforth use the term \textit{perturbation} to refer to an error of magnitude $\epsilon$ in the entries of $A$ (i.e. $A_{i,j}+\epsilon_{i,j}$).

The results we obtain are organized as follows:
Section \ref{section:SignSwitches} is dedicated to investigating the case when a single entry of $A$ is perturbed. 
We begin in Section \ref{section:PerturbSingleEntry} by applying the well-known Sherman-Morrison formula for the inverse of a rank 1 perturbation of a matrix. This allows us obtain an inequality that determines when the sign of the $(i,j)$ entry of the inverse of the perturbed matrix differs from the sign of the $(i,j)$ entry of the inverse of the unperturbed matrix. Using this, we calculate explicitly the probability of a sign switch in the $(i,j)$ entry of the inverse when the perturbation is taken according to some arbitrary distribution. 
In Section \ref{section:LimitingValue}, we then study the limiting behavior of the total number of sign switches in the inverse of the perturbed matrix as the perturbation grows to infinity and explicitly quantify how large the perturbation must be to reach this limiting number of sign switches. We also numerically determine the expected fraction of sign switches using independent Gaussian matrices and compare this with the two motivating motif examples. In Section \ref{section:Norm}, we investigate how perturbing a single entry of a given matrix affects the norm of the inverse, relative to the norm of the inverse of the unperturbed matrix. We use the entry-wise (vectorized) 1-norm and the (spectral) 2-norm to investigate these characterizations of quantitative sensitivity.

In Section \ref{section:PerturbManyEntries}, we investigate the case of perturbing multiple entries of $A$. Utilizing an iterative application of the Sherman-Morison formula, we show how to obtain a system of inequalities that describes when sign switches will occur in the inverse of the perturbed matrix. Looking at the region of ``perturbation space'' where no sign switch occurs gives an indication of the ``sign sensitivity'' of the matrix, and we define two indices reflecting this quantity.

In Section \ref{section:TriDiag}, we derive conditions on a tridiagonal matrix that ensure that the sign pattern of the inverse of the matrix depends only on the sign pattern of the original matrix. This defines a class of matrices in which arbitrary perturbations (as long as they don't change the sign of the original matrix) will \textit{not} cause the inverse of the perturbed matrix to have a different sign pattern than that of the inverse of the unperturbed matrix.  The trophic chain (TC) motif provides an illustrative special case.

Finally, in Section \ref{section:IGPExample}, we provide an indication of why the IGP motif exhibits a high proclivity for switching signs under any set of perturbations, as is demonstrated in Section \ref{section:PerturbManyEntries}. We accomplish this by illustrating a way to decompose this matrix into a sum of terms that can be individually analyzed using the techniques developed in Sections \ref{section:SignSwitches} and \ref{section:TriDiag}.

Throughout this paper, unless otherwise stated, we assume that $A$ is invertible and that for $A_{i,j}^{-1}$, the $(i,j)$ entry of $A^{-1}$, we have $-A^{-1}_{i,j}\neq 0$ for each $i$ and $j$.  When we consider a perturbation by another matrix $B$, we assume that $A+B$ is invertible. We typically think of $A$ as a square, $n\times n$ real matrix. For notational simplicity, we consider $A^{-1}$ and not $-A^{-1}$ as this choice does not effect the determination of a sign switch.

\section{Motivating examples}
{ Following \cite{Takimoto:2007ek,Novak:2016zr}, consider a four-species system (Figure \ref{fig:motifs}b) described by the following set of differential equations:
\begin{align} \label{eqn:IGPmotif}
f_1=\frac{dN_1}{dt} &= I + (r_1 + a_{1,2} N_2)N_1\\ \notag
f_2=\frac{dN_2}{dt} &= (r_2 + a_{2,1} N_1+ a_{2,2}N_2 + a_{2,3}N_3 + a_{2,4} N_4)N_2\\ \notag
f_3=\frac{dN_3}{dt} &= (r_3 + a_{3,2} N_2 + a_{3,3}N_3 + a_{3,4} N_4)N_3\\ \notag
f_4=\frac{dN_4}{dt} &= (r_4 + a_{4,2} N_2 + a_{4,3}N_3 + a_{4,4} N_4)N_4,
\end{align}
with $a_{i,j}, r_i, I \in \mathbb{R}$ and where at time $t\in \mathbb{R}$ the abundance of species $i$ is given by $N_i(t)\in \mathbb{R}$. The $a_{i,j}$, $r_i$, and $I$ parameters respectively represent the per capita strengths of the species' interactions, the species' intrinsic per capita growth (death) rates, and a density-independent influx to the basal species.

The direct effects between each pair of species can be given in terms of this system's Jacobian with $A_{ij}=\tfrac{\partial f_i}{\partial N_j}$:}
\begin{align}
\label{eqn:IGPmotifA}
A=\left(
\begin{array}{cccc}
 A_{1,1} & A_{1,2} & 0 & 0 \\
 A_{2,1} & A_{2,2} & A_{2,3} & A_{2,4} \\
 0 & A_{3,2} & A_{3,3} & A_{3,4} \\
 0 & A_{4,2} & A_{4,3} & A_{4,4} \\
\end{array}
\right).
\end{align}
The structural form of the equations given in \eqref{eqn:IGPmotif} and summarized in the matrix $A$ in \eqref{eqn:IGPmotifA} (with variable $A_{i,j}$'s) is referred to as the intraguild predation motif (or IGP motif). A realization of this motif is given by fixing values for the $A_{i,j}$. Given fixed values for all $A_{i,j}$, the sign pattern of $-A^{-1}$ is equivalent to the sign pattern of ${\rm adj}(-A)$. 

Given fixed values for the $A_{i,j}$'s, we wish to study the change in $-A^{-1}$ as a function of perturbing (adding error to) the $A_{i,j}$'s: replacing $A_{i,j}$ with $A_{i,j} + \epsilon$ representing the uncertainty in the value of $A_{i,j}$. For example, using a realization of the IGP motif given by
{\footnotesize \begin{align}
\label{IGPMatrix}
\IGP=\left(
\begin{array}{cccc}
 -0.237 & -1 & 0 & 0 \\
 0.1 & -0.015 & -1 & -1 \\
 0 & 0.1 & -0.015 & -1 \\
 0 & 0.045 & 0.1 & -0.015 \\
\end{array}\right)
\end{align}}
\noindent
implies that
{\footnotesize
\begin{align}
-{\IGP}^{-1}=\left(
\begin{array}{cccc}
 12.05 & 38.56 & -44.24 & 378.95 \\
 -3.86 & -9.14 & 10.49 & -89.81 \\
 1.67 & 3.97 & -4.70 & 48.96 \\
 -0.41 & -0.97 & 0.12 & -9.72 \\
\end{array}
\right),
\end{align}}
\noindent
reflecting the response of the species in row $i$ to a press perturbation of the species in column $j$. For comparison, perturbing the $(4,2)$ entry of $\IGP$ by an error of magnitude 1 leads to 
{\footnotesize \begin{align*}
-A_\epsilon ^{-1}=\left(
\begin{array}{cccc}
 -4.4 & -0.43 & 0.5 & -4.27 \\
 0.04 & 0.1 & -0.12 & 1.01 \\
 -0.45 & -1.07 & 1.08 & -0.55 \\
 0.01 & 0.03 & -1.03 & 0.11 \\
\end{array}
\right),
\end{align*}}%

\noindent
demonstrating that such a perturbation leads to a sign change in each entry of the perturbed system (and hence a different qualitative prediction of each component of the system).

For illustrative comparisons to the IGP motif, we use the following realization of the trophic chain motif (Figure \ref{fig:motifs}a): 
{\footnotesize\begin{align}
\label{TriMatrix}
\Tri=\left(
\begin{array}{cccc}
 -0.237 & -1 & 0 & 0 \\
 0.1 & -0.015 & -1 & 0 \\
 0 & 0.1 & -0.015 & -1 \\
 0 & 0 & 0.1 & -0.015 \\
\end{array}
\right).
\end{align}}
Note that $\Tri$ is obtained from $\IGP$ by setting $A_{2,4}=A_{4,2}=0$.

\section{Sign Switches of the Inverse of an Arbitrary Matrix Under Perturbation}
\label{section:SignSwitches}
We begin by introducing our technique for determining the number of sign switches that perturbations in the entries of $A$ will incur.  To compute the inverse of the sum of two matrices $(A+B)^{-1}$ when the inverse of one, say $A^{-1}$, is known, we use a result of \cite{miller1981inverse}. Here, we think of $B$ as a perturbation of $A$. 
\begin{theorem}[Lemma 1 of\cite{miller1981inverse}]
\label{thm:Rank1Pert}
Let $A$ and $A+B$ be nonsingular matrices where $B$ is rank one. Let $g={\rm tr}\left( BA^{-1} \right) = \sum_{i,j} B_{j,i} A^{-1}_{i,j}$. Then $g\neq -1$ and
\begin{align}
\label{eqn:MillerTheorem}
(A+B)^{-1} = A^{-1}-\frac{1}{1+g} A^{-1}BA^{-1}.
\end{align}
\end{theorem}
This formula is also known as the Sherman-Morrison formula.

\subsection{Perturb Single Entry}
\label{section:PerturbSingleEntry}
We apply Theorem \ref{thm:Rank1Pert} in the case of perturbing a single entry of the matrix $A$. That is, for $\delta_{k,l}$ being the matrix (of same size as $A$) of zeros save a single 1 in the $(k,l)$ entry and zero otherwise, $B=\epsilon \delta_{k,l}$ where $\epsilon$ is the magnitude of the perturbation. Applying Theorem \ref{thm:Rank1Pert}, we have the $(i,j)$ entry of $(A+\epsilon \delta_{k,l})^{-1}$ given by:
\begin{align}
\label{eqn:perturbsingleentry}
(A+\epsilon \delta_{k,l})^{-1}_{i,j} = A_{i,j}^{-1} - \frac{\epsilon A^{-1}_{i,k} A^{-1}_{l,j}}{1+\epsilon A^{-1}_{l,k}}.
\end{align}
Upon dividing by $A^{-1}_{i,j}$, this leads to the following Lemma:
\begin{lemma}
\label{lemma:PerturbSingle}
Given a fixed invertible matrix $A$ and an invertible perturbation $A+\epsilon \delta_{k,l}$ such that for each $i,j$, $A^{-1}_{i,j}\neq 0$ and that $A_{l,k}^{-1}\neq \frac{-1}{\epsilon}$, the $(i,j)^{\rm th}$ entry of $A^{-1}$ will have a different sign from the $(i,j)^{\rm th}$ entry of $(A+\epsilon \delta_{k,l})^{-1}$ if and only if
\begin{align}
\label{eq:PerturbSingleformula}
1 - \frac{\epsilon A^{-1}_{i,k} A^{-1}_{l,j}}{A_{i,j}^{-1}\left(1+\epsilon A^{-1}_{l,k}\right)} < 0 .
\end{align}
\end{lemma}
\noindent
In words, this means that an error to the direct effect of species $l$ on species $k$ will cause a qualitatively incorrect prediction to be made for the net effect of any $j^{th}$ species on any $i^{th}$ species if and only if the ratio of: 1. the product of the error and the net effects of $k$ on $i$ and of $j$ on $l$ and 2. the net effect of $j$ on $i$ times one plus the product of the error and the net effect of $k$ on $l$, is greater than 1.

If $\epsilon$ is not a fixed quantity but rather drawn according to some distribution, then Lemma \ref{lemma:PerturbSingle} can be utilized to calculate the probability of a sign switch in the $(i,j)^{\rm th}$ entry of $(A+\epsilon \delta_{k,l})^{-1}$. For notational simplicity, let $C=A^{-1}$, then assuming that $\epsilon$ is drawn from a uniform distribution on $[0,1]$, and, for example, assuming that $C_{l,k}<-1$ and $C_{i,j}<0$, we have the probability that $A^{-1}_{i,j}$ differs in sign from $(A+\epsilon \delta_{k,l})^{-1}_{i,j}$ given by $1-p$ where $p$ has the value:
\begin{align*}
p=\left\{
\begin{array}{ll}
 1+\frac{1}{C_{l,k}}, & {\rm if}\ \left(C_{i,k}<0\lor C_{l,j}>0\right)\land \left(C_{i,k}>0\lor C_{l,j}\leq 0\right)\land \frac{C_{i,k} C_{l,j}}{C_{i,j}}\leq C_{l,k}+1 \\
- \frac{C_{i,j}}{C_{i,k} C_{l,j}+C_{i,j} C_{l,k}}-\frac{1}{C_{l,k}}, & {\rm if}\ \left(C_{i,k}>0\land C_{l,j}\leq 0\right)\lor \left(C_{i,k}<0\land C_{l,j}>0\right) \\
-\frac{C_{i,j}}{C_{i,j} C_{l,k}+C_{i,k} C_{l,j}}+\frac{1}{C_{l,k}}, & {\rm Otherwise.} \\
\end{array}
\right.
\end{align*}
Observe that this formula is explicitly given in terms of the entries of $C=A^{-1}$.

Furthermore, if $\epsilon$ is a random variable, one can use the expression in equation \eqref{eq:PerturbSingleformula} to define another random variable indicating if a sign switch has occurred:
\begin{align}
\label{eq:FuncRandVar}
\mathbbm{1}_{{\rm switch}(A,\epsilon,i,j,k,l)} = \left\{\begin{array}{lr} 1, &{\rm if}\ 1 - \frac{\epsilon A^{-1}_{i,k} A^{-1}_{l,j}}{A_{i,j}^{-1}\left(1+\epsilon A^{-1}_{l,k}\right)} < 0 \\
0, &{\rm if}\ 1 - \frac{\epsilon A^{-1}_{i,k} A^{-1}_{l,j}}{A_{i,j}^{-1}\left(1+\epsilon A^{-1}_{l,k}\right)} > 0.
\end{array} \right.
\end{align}
Summing over $i$ and $j$ will return the number of sign switches that have occurred in $A^{-1}$ when perturbing the $(k,l)$ entry of $A$ by magnitude $\epsilon$:
\begin{align}
\label{eq:NumSwitch}
{\rm \NS}(A,\epsilon,k,l) = \sum_{i,j} \mathbbm{1}_{{\rm switch}(A,\epsilon,i,j,k,l)}.
\end{align}
Note that the number of switches ${\rm \NS}(A,\epsilon,k,l)$ is a deterministic (non-random) function of $A$ and a fixed $\epsilon$. For fixed $k$, $l$, and $A$, the value of ${\rm \NS}(A,\epsilon,k,l)$ is an integer. Hence, for random $\epsilon$, calculating the expected number of sign switches can be accomplished as follows: Let $f_\epsilon$ be the probability density function of the random variable $\epsilon$, then the expected total number of sign switches in $A^{-1}$ when perturbing the $(k,l)$ entry by a magnitude given by (the random variable) $\epsilon$ is given by:
\begin{align}
\label{eq:ExpectedNumberSingle}
\mathbbm{E}({\rm \NS}(A,\epsilon,k,l)) = \int {\rm \NS}(A,x,k,l)\ f_\epsilon(x)\ dx.
\end{align}

\begin{remark}
Picturing the number of switches ${\rm \NS}(A,\epsilon,k,l)$ as a function of $\epsilon$ for various $k$ and $l$ reveals which entries $(k,l)$ cause the most sign switches in $A^{-1}$ when they are perturbed. For example, the number of expected sign switches as a function of $\epsilon$ for the matrix $\IGP$ from equation \eqref{IGPMatrix} is illustrated in Figure \ref{fig:NumSwitchEx}. Also shown in Figure \ref{fig:NumSwitchEx} is a illustrative distribution for $\epsilon$, chosen not to change the sign of the $(k,l)$ entry of $A$: $\epsilon \sim {\rm sign}(A_{k,l}){\rm Exp}(1)-A_{k,l}$. 

It follows that the expected fraction of sign switches, over all $n^2$ entries in the matrix, is given by:
\begin{align}
\label{eqn:ExpectedFraction}
\frac{1}{n^2}\mathbbm{E}\left({\rm \NS}\left(A,\epsilon,k,l\right)\right)
\end{align} 
For the example matrix $\IGP$, this leads to:
\begin{align*}
\frac{1}{4^2}\mathbbm{E}\left({\rm \NS}\left(\IGP,\epsilon,4,3\right)\right) &= 0.18\\
\frac{1}{4^2}\mathbbm{E}\left({\rm \NS}\left(\IGP,\epsilon,2,3\right)\right) &= 0.34.
\end{align*}
\noindent
This means, for example, that perturbing the $(2,3)$ entry in $\IGP$ according to the aforementioned distribution will cause an average of $34\%$ of qualitatively incorrect predictions.

In Figure \ref{fig:QualSens} we include heat maps of the percent of expected sign switches $100\% \times \frac{1}{4^2}\mathbbm{E}({\rm \NS}(A,\epsilon,k,l))$ over all $k$ and $l$ for each example matrix. The distribution over which the expectation is taken is given by $\epsilon \sim {\rm sign}(A_{k,l}){\rm Exp}(1)-A_{k,l}$.  Figure \ref{fig:QualSens}a) illustrates that the qualitative dynamics of the TC motif realization are entirely sign insensitive to any quantitative uncertainty in $\Tri$.  Figure \ref{fig:QualSens}b) illustrates, for example, how the qualitative dynamics of the IGP motif realization are sign insensitive to uncertainty in top-down direct effect of the top-predator (species 4) on the shared resource (species 2) (i.e. $A_{2,4}$), but sign sensitive to uncertainty in the reciprocal bottom-up effect of the shared resource on the top-predator (i.e. $A_{4,2}$).

\begin{figure}[!h]%
\begin{center}
\includegraphics[width=6.25in,trim={0 0 0 0in},clip]{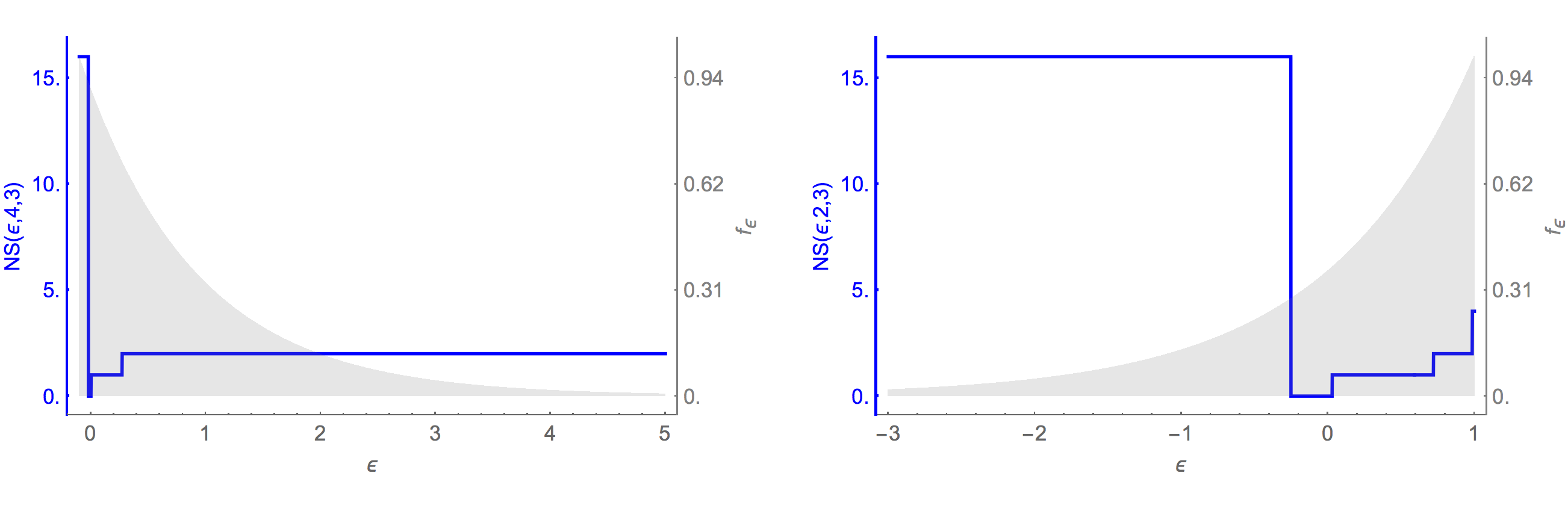}%
\end{center}
\caption{{The number of sign switches as a function of the error in the ($k,l$) entry of $\IGP$ for $(k,l)=(4,3)$ and $(k,l)=(2,3)$. Overlain on these is the probability density function $f_\epsilon$ for $\epsilon \sim {\rm sign}(A_{k,l}){\rm Exp}(1)-A_{k,l}$, describing an illustrative and empirically likely declining probability of making large errors to $A_{k,l}$.  Such a distribution of errors for the two entries respectively results in getting 18\% and 34\% of predictions in $-A^{-1}$ qualitatively incorrect.} }
\label{fig:NumSwitchEx}%
\end{figure}

\begin{figure}[!h]%
\begin{center}
\includegraphics[width=3.25in,trim={0 0 0 0in},clip]{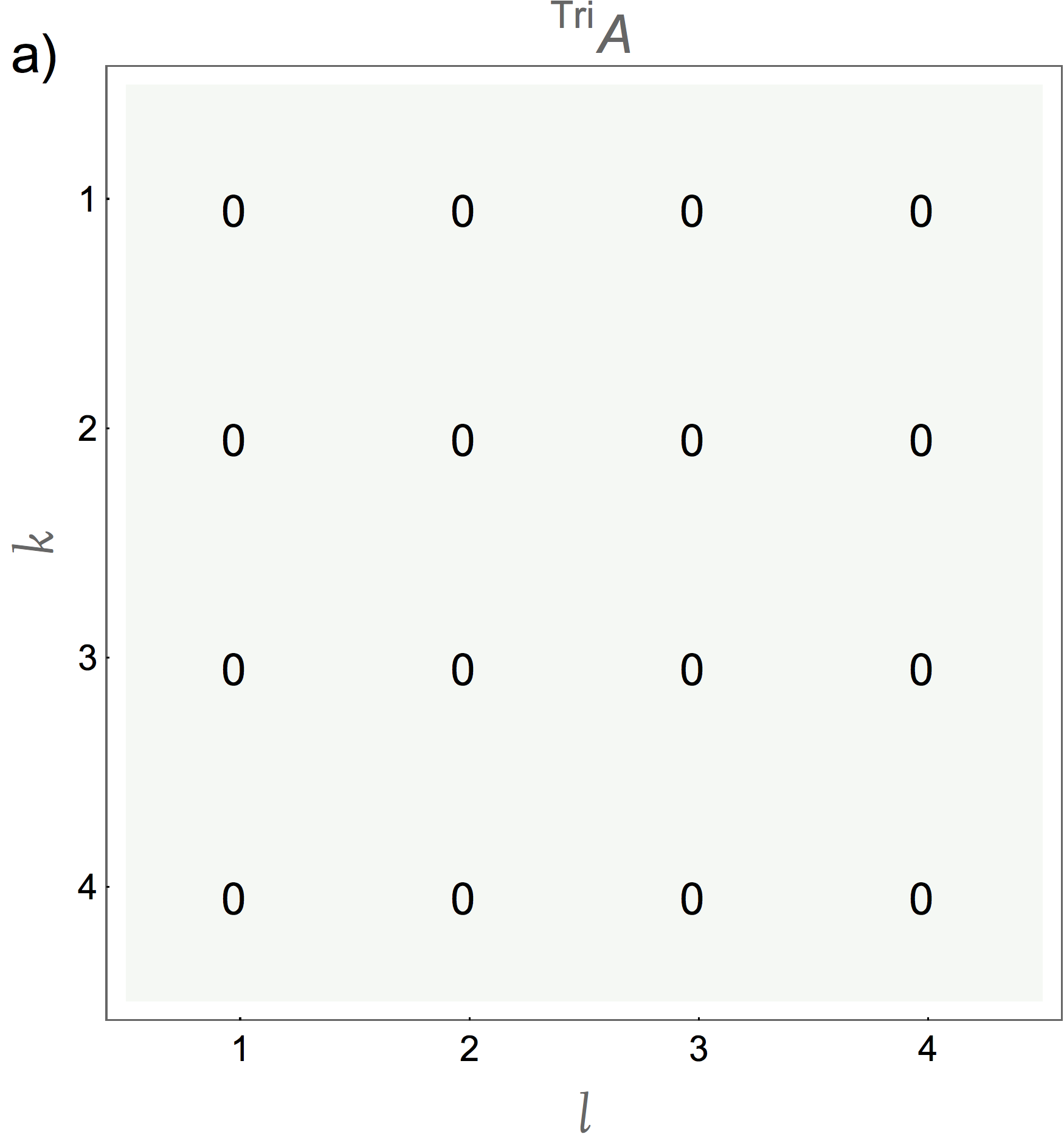}%
\includegraphics[width=3.25in,trim={0 0 0 0in},clip]{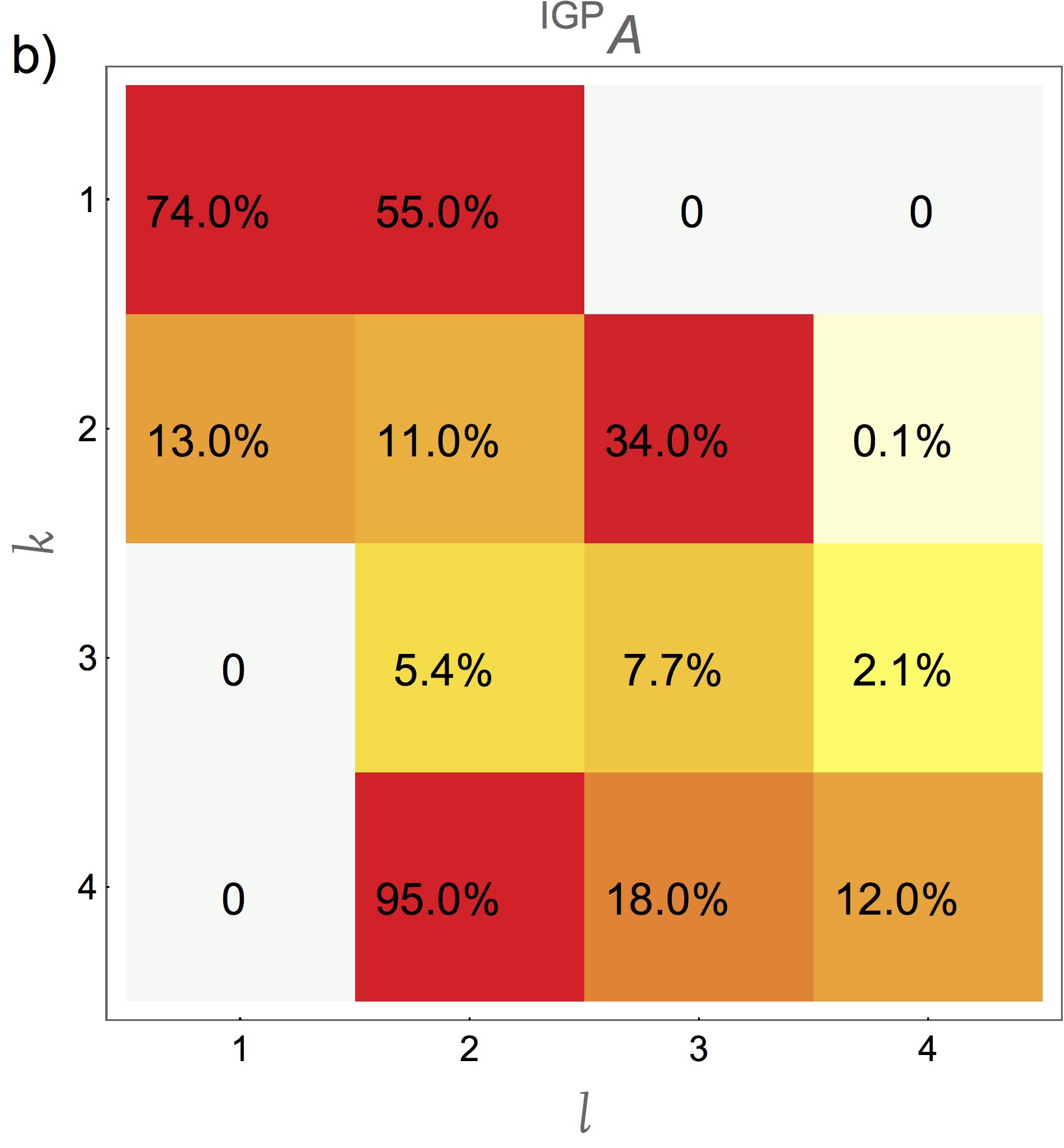}
\end{center}
\caption{Heat maps for the average percent of sign switches (defined in equation \eqref{eqn:ExpectedFraction}) induced by perturbation of each $(k,l)$ entry. Perturbations are given by the distribution $\epsilon \sim {\rm sign}(A_{k,l}){\rm Exp}(1)-A_{k,l}$. a) The trophic chain motif. b) The intraguild predation motif. Note that the trophic chain undergoes no sign switches when perturbed (see Theorem \ref{thm:Tinverse} in Section \ref{section:TriDiag} for an explanation of why) while the intraguild predation motif on average over the perturbed $k$ and $l$ experiences $27.2\%$ of entries (that is, 4.4 entries) being qualitatively incorrect. Compare this to the quantitative sensitivity given in Figure \ref{fig:QuantSens}.}
\label{fig:QualSens}%
\end{figure}

\end{remark}
\subsection{Limiting value of \NS}
\label{section:LimitingValue}
In this section, we focus on describing the behavior of the number of sign switches ${\rm \NS}(A,\epsilon,k,l)$ as $\epsilon \rightarrow \pm \infty$ and quantify how large $\epsilon$ must be to reach this limiting value. Before describing the limiting behavior, we prove a Lemma that gives an expression for ${\rm \NS}(A,\epsilon,k,l)$. In the following, we often use the $k$-minors of a matrix, defined as follows.
\begin{definition}[$k$-minors]
For an $n\times n$ matrix $A$, an integer $k<n$, unique integers $1<i_1,i_2,\dots,i_k\leq n$, and unique integers $1\leq j_1,j_2,\dots,j_k\leq n$, let $A_{[i_1\cdots i_k,j_1\cdots j_k]}$ be the sub-matrix of $A$ obtained by deleting rows $i_1$ through $i_k$ and columns $j_1$ through $j_k$. By a $k$-minor, we mean a determinant of the form $\det\left(A_{[i_1\cdots i_k,j_1\cdots j_k]}\right)$. By a minor, we mean a $1$-minor (i.e. determinant after deleting one row and one column).
\end{definition}

{We first note that for certain values of $i$ and $j$, the inequality in equation \eqref{eq:PerturbSingleformula} reduces to an expression not involving $i$ or $j$. As a consequence, depending only on the perturbation magnitude $\epsilon$ and the size of the entry $A^{-1}_{l,k}$, perturbing the entry in the $k^{\rm th}$ row and $l^{\rm th}$ column of a matrix $A$ causes the signs of \textit{all} entries of the $l^{\rm th}$ \textit{row} and $k^{\rm th}$ \textit{column} of the inverse to either match or not match the signs of the corresponding elements in the unperturbed matrix.
}

\begin{lemma}
\label{lemma:lkminor}
For an invertible $n\times n$ matrix $A$ such that no entry of $A^{-1}$ is zero, for $1\leq i,j,k,l\leq n$ such that $i=l$, or $j=k$, or $\det(A_{[ki,lj]})=0$, then the expression
\begin{align}
1-\frac{\epsilon A^{-1}_{i,k} A^{-1}_{l,j}}{A^{-1}_{i,j}\left(1+\epsilon A^{-1}_{l,k}\right)}<0
\end{align}
is true if and only if the logical expression
\begin{align}
(\epsilon<0\ {\rm and}\ A^{-1}_{l,k}>\frac{-1}{\epsilon})\ {\rm or}\ (\epsilon>0\ {\rm and}\ A^{-1}_{l,k}<\frac{-1}{\epsilon})
\end{align}
is true.
\end{lemma}
\noindent
This result implies that once an error associated with the direct effect of species $l$ on species $k$ causes a qualitatively incorrect prediction to be made for \textit{any one} of the network's net effects that either emanate from species $k$ or affect species $l$, then \textit{all} predictions for the net effects emanating from $k$ and affecting $l$ will be qualitatively incorrect (and hence can be corrected by making the opposite qualitative prediction).

\begin{proof}
The proof of this Lemma is via computation utilizing Lemma \ref{lemma:PerturbSingle}.
 Namely, when $i=l$,
\begin{align}
1-\frac{\epsilon A_{l,k}^{-1}A^{-1}_{l,j}}{A^{-1}_{l,j}\left(1+\epsilon A^{-1}_{l,k}\right)} = 1- \frac{\epsilon A^{-1}_{l,k}}{1+\epsilon A^{-1}_{l,k}}.
\end{align}
Note that $1- \frac{\epsilon A^{-1}_{l,k}}{1+\epsilon A^{-1}_{l,k}}<0$ if and only if either $\epsilon<0\ {\rm and}\ A^{-1}_{l,k}>\frac{-1}{\epsilon}$ or $\epsilon>0\ {\rm and}\ A^{-1}_{l,k}<\frac{-1}{\epsilon}$.
A similar calculation takes care of the case when $j=k$. In the case that $\det(A_{[ki,lj]})=0$, by Theorem 2.5.2 in \cite{prasolov1994problems}, $\det(A_{[ki,lj]})=0$ if and only if $A^{-1}_{i,k}A^{-1}_{l,j} = A^{-1}_{i,j}A^{-1}_{l,k}$, hence
\begin{align}
1-\frac{\epsilon A^{-1}_{i,k} A^{-1}_{l,j}}{A^{-1}_{i,j}\left(1+\epsilon A^{-1}_{l,k}\right)} = 1-\frac{\epsilon A^{-1}_{i,j} A^{-1}_{l,k}}{A^{-1}_{i,j}\left(1+\epsilon A^{-1}_{l,k}\right)} = 1-\frac{\epsilon A^{-1}_{l,k}}{1+\epsilon A^{-1}_{l,k}}
\end{align}
which, as before, is less than zero if and only if either $\epsilon<0\ {\rm and}\ A^{-1}_{l,k}>\frac{-1}{\epsilon}$ or $\epsilon>0\ {\rm and}\ A^{-1}_{l,k}~<~\frac{-1}{\epsilon}$.
\end{proof}

This Lemma allows us to obtain an expression for ${\rm \NS}(A,\epsilon,k,l)$ by separating the terms where  $\det(A_{[ki,lj]})=0$. In the following, let $N(k,l)=\left|\left\{1\leq i,j\leq n:\ i\neq l, j\neq k, \det(A_{[ki,lj]})=0\right\}\right|$ be the number of $2$-minors equal to zero involving the $k^{\rm th}$ row and $l^{\rm th}$ column (with $i\neq l$ and $j\neq k$).
\begin{theorem}
\label{thm:NumSwitchReduction}
For an invertible $n\times n$ matrix $A$ such that no entry of $A^{-1}$ is zero, and for $1\leq k,l\leq n$, and $\epsilon \in \mathbbm{R}$,
\begin{align}
\begin{split}
{\rm \NS}(A,\epsilon,k,l) &= \sum_{\substack{i\neq l\\ j\neq k\\ \det(A_{[ki,lj]})\neq 0}} \mathbbm{1}_{\left\{ \frac{\epsilon A^{-1}_{i,k} A^{-1}_{l,j}}{A^{-1}_{i,j} \left(1+\epsilon A^{-1}_{l,k}\right)}>1\right\}}\\ &+ (2n-1+N(k,l)) \mathbbm{1}_{\left\{ \left(\epsilon<0\ \land\ A^{-1}_{l,k}>\frac{-1}{\epsilon}\right)\ \lor\ \left(\epsilon>0\ \land\ A^{-1}_{l,k}<\frac{-1}{\epsilon}\right)\right\}}.
\end{split}
\end{align}
\end{theorem} \QEDB

This Theorem allows us to determine the limiting behavior of the number of switches ${\rm \NS}(A,\epsilon,k,l)$. In particular, after a certain point, increasing the magnitude of the perturbation $\epsilon$ no longer causes a sign switch in the inverse of the perturbed matrix when compared to the original matrix.

\begin{corollary}
\label{cor:limitingbehavior}
For an invertible $n\times n$ matrix $A$ such that no entry of $A^{-1}$ is zero, and for $1\leq k,l\leq n$, if $\epsilon_1, \epsilon_2\in \mathbbm{R}$ such that
$$
|\epsilon_1|,|\epsilon_2| > \max_{\substack{i\neq l\\ j\neq k\\ \det(A_{[ki,lj]})\neq 0}} \left|\frac{A^{-1}_{i,j}}{A^{-1}_{i,k}A^{-1}_{l,j}-A^{-1}_{i,j}A^{-1}_{l,k}}\right|
$$
then 
$$
{\rm \NS}(A,\epsilon_1,k,l) = {\rm \NS}(A,\epsilon_2,k,l).
$$
In particular, when ${\rm sign}(\epsilon_1)={\rm sign}(A_{k,l})$ (i.e. the error changes the magnitude of $A_{k,l}$ but not its sign), we have
\begin{align}
\label{eqn:limitingbehavior}
{\rm \NS}(A,\epsilon_1,k,l) = \sum_{\substack{i\neq l,j\neq k: \det(A_{[ki,lj]})\neq 0}} \mathbbm{1}_{\left\{ \frac{ A^{-1}_{i,k} A^{-1}_{l,j}}{A^{-1}_{i,j} A^{-1}_{l,k}}>1\right\}} + (2n-1+N(k,l)) \mathbbm{1}_{\left\{{\rm sign}\left(A^{-1}_{l,k}\right) \neq {\rm sign}(A_{k,l})\right\}}.
\end{align}
\end{corollary}
\begin{proof}
First, note that as a function of $\epsilon$, the quantity $\frac{\epsilon A^{-1}_{i,k} A^{-1}_{l,j}}{A^{-1}_{i,j}(1+\epsilon A^{-1}_{l,k})}$ is monotonic (depending on the sign of $\frac{A^{-1}_{i,k}A^{-1}_{l,j}}{A^{-1}_{i,j}}$) with a discontinuity at $\epsilon = -\frac{1}{A^{-1}_{l,k}}$. Furthermore, $\frac{\epsilon A^{-1}_{i,k} A^{-1}_{l,j}}{A^{-1}_{i,j}(1+\epsilon A^{-1}_{l,k})}=1$ if and only if $\epsilon = \frac{1}{\frac{A^{-1}_{i,k}A^{-1}_{l,j}}{A^{-1}_{i,j}}-A^{-1}_{l,k}}=\frac{A^{-1}_{i,j}}{A^{-1}_{i,k}A^{-1}_{l,j}-A^{-1}_{l,k}A^{-1}_{i,j}}$. Now we know that  
\begin{align}
\lim_{\epsilon \rightarrow \pm \infty} \frac{\epsilon A^{-1}_{i,k} A^{-1}_{l,j}}{A^{-1}_{i,j}\left(1+\epsilon A^{-1}_{l,k}\right)} = \frac{A^{-1}_{i,k} A^{-1}_{l,j}}{A^{-1}_{i,j}A^{-1}_{l,k}},
\end{align}
hence, for $i\neq l, j\neq k$ and such that $\det(A_{[ki,lj]})\neq 0$, we have that as long as 
$$
|\epsilon_1|>\max_{\substack{i\neq l\\ j\neq k\\ \det(A_{[ki,lj]})\neq 0}} \left|\frac{A^{-1}_{i,j}}{A^{-1}_{i,k}A^{-1}_{l,j}-A^{-1}_{i,j}A^{-1}_{l,k}}\right|,
$$
we have that
\begin{align}
\frac{A^{-1}_{i,k} A^{-1}_{l,j}}{A^{-1}_{i,j}A^{-1}_{l,k}}>1\quad {\rm if\ and\ only\ if}\quad \frac{\epsilon_1 A^{-1}_{i,k} A^{-1}_{l,j}}{A^{-1}_{i,j}\left(1+\epsilon_1 A^{-1}_{l,k}\right)}>1.
\end{align}

Combining this with the definition given in line \eqref{eq:NumSwitch} proves the first part of the Corollary. The second part of the Corollary is proved by applying this observation to Theorem \ref{thm:NumSwitchReduction} and noting that as $|\epsilon|\rightarrow \infty$, the expression $(\epsilon<0\ \land\ A^{-1}_{l,k}>\frac{-1}{\epsilon})\ \lor\ (\epsilon>0\ \land\ A^{-1}_{l,k}<\frac{-1}{\epsilon})$ is true if and only if ${\rm sign}\left(A^{-1}_{l,k}\right) \neq {\rm sign}(\epsilon)={\rm sign}(A_{k,l})$.
\end{proof}

A useful application of Corollary \ref{cor:limitingbehavior} is when the matrix $A$ has all its entries drawn independently from a continuous distribution, such as the standard normal distribution (i.e. each entry of $A$ is drawn from an independent standard normal random variable). In the limit, the expected fraction of entries that experience a sign switch has an especially compact representation.

\begin{corollary}
\label{cor:Gauss}
For an $n\times n$ matrix $A$ with entries drawn independently from a continuous distribution (such as independent standard normal random variables), the asymptotic expected fraction of sign switches is given by
\begin{align}
\label{eqn:expectednumswitch}
\lim_{n \rightarrow \infty} \lim_{|\epsilon|\rightarrow \infty} \mathbbm{E}_A\left({\rm \NS}(A,\epsilon,k,l)\right)/n^2 = \lim_{n \rightarrow \infty} \frac{1}{n^2} \sum_{i,j=1}^n \mathbbm{P}\left(\frac{A^{-1}_{i,k} A^{-1}_{l,j}}{A^{-1}_{i,j} A^{-1}_{l,k}}>1\right). 
\end{align}
\end{corollary}
{Thus, for a matrix with size approaching infinity with all entries drawn independently from a standard normal distribution, the proportion of qualitatively incorrect predictions can be expressed in a form that eases theoretical and computational calculations.}
\begin{proof}
For fixed $k$ and $l$, the set of matrices $A$ such that there exist $i,j$ such that $\det(A_{[ki,lj]})=0$ has positive codimension in the set of all real, invertible $n\times n$ matrices and hence has probability zero for the continuous distribution under consideration. Hence the quantity $(2n-1+N(k,l))$ in line \eqref{eqn:limitingbehavior} is equal to $2n-1$ corresponding to the cases $i=l$ or $j=k$. However, since $\frac{2n-1}{n^2}\rightarrow 0$ as $n\rightarrow \infty$, 
applying the expectation to Corollary \ref{cor:limitingbehavior} leads to the result.
\end{proof}

\begin{remark}
Unfortunately, it appears not to be straightforward to exactly compute the quantity $\mathbbm{P}\left(\frac{A^{-1}_{i,k} A^{-1}_{l,j}}{A^{-1}_{i,j} A^{-1}_{l,k}}>1\right)$ for each $i,j$. Nevertheless, the expression in equation \eqref{eqn:expectednumswitch} does give a convenient way to sample the limiting expected fraction of switches (as the right hand side is free from the perturbation value $\epsilon$). To demonstrate, we fixed $n=200$ and $k=25$, $l=70$ and computed the right hand side of equation \eqref{eqn:expectednumswitch} for 7,000 draws of matrices with independent standard normal entries. The resulting distribution is show in Figure \ref{fig:AveNumSwitch} and does not appear to depend significantly on the chosen $k$ and $l$. Interestingly, the average percent of sign switches in this computation was approximately 33\%, and the resulting distribution appears to be similar to a beta distribution (shown in Figure \ref{fig:AveNumSwitch} with a blue line) indicating that further simplifications of \eqref{eqn:expectednumswitch} may be possible.

\begin{figure}[!h]%
\begin{center}
\includegraphics[width=5.25in,trim={0 0 0 0in},clip]{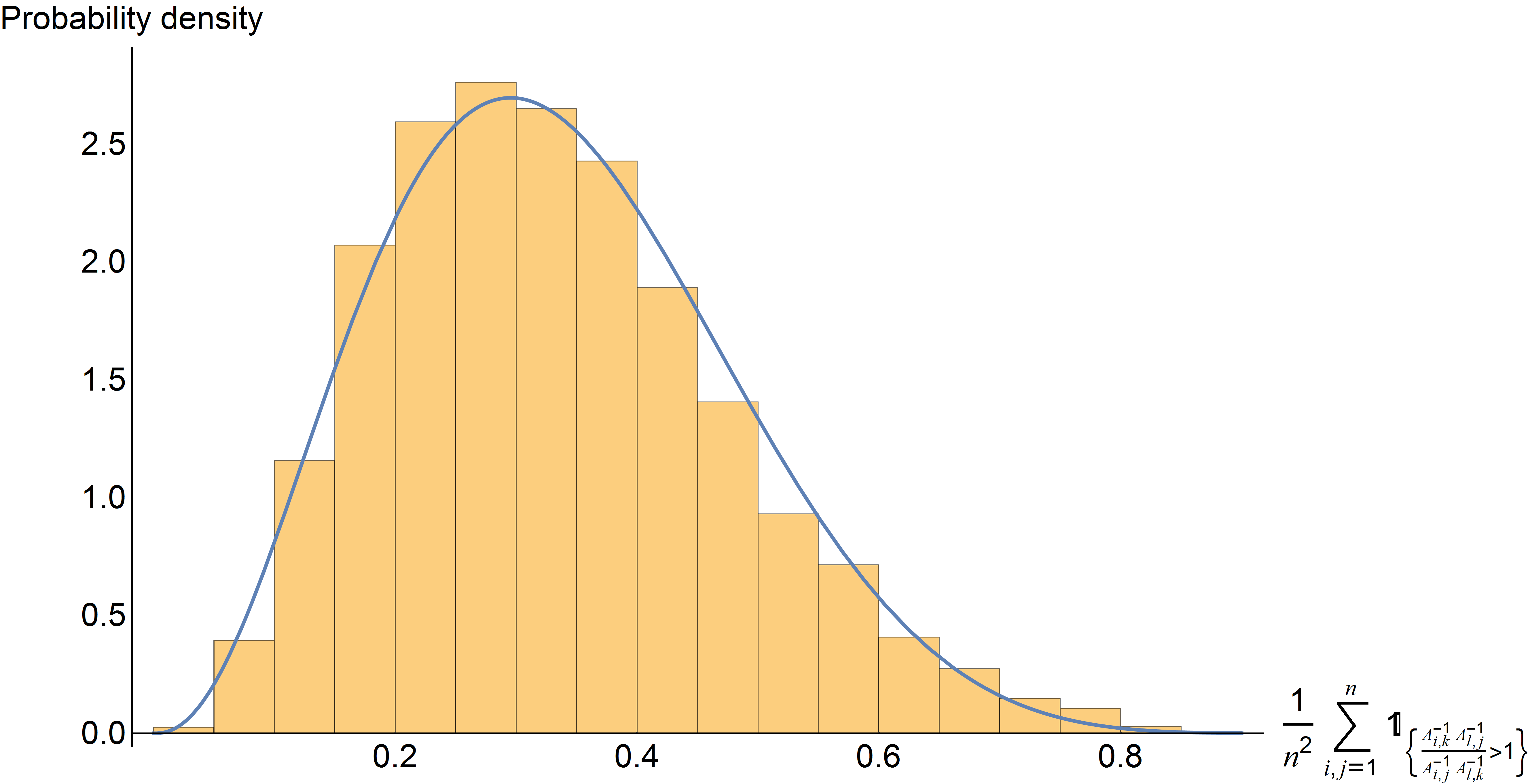}%
\end{center}
\caption{Histogram of $\frac{1}{n^2} {\rm \NS}(A,\epsilon,k,l)$ obtained by sampling $A$ with independent standard normal entries and evaluating the right hand side of equation \eqref{eqn:expectednumswitch} 7,000 times using $n~=~200$, $k=25$, and $l=70$. The probability density function for a ${\rm Beta}(3.4088,6.7448)$ distribution is overlaid this histogram as a blue line.}
\label{fig:AveNumSwitch}%
\end{figure}

In comparison, even for small $n$ such as when $n=4$, using Corollary \ref{cor:limitingbehavior}, the mean (averaged over $k$ and $l$ and 5,000 draws of a entry-wise independent standard normal distribution) expectation of signs switches is give by,
$$
\lim_{|\epsilon|\rightarrow \infty} \frac{1}{4^2}\sum_{k,l}\mathbbm{E}_A\left({\rm \NS}(A,\epsilon,k,l)\right) \approx 0.33.
$$
We compare this value to our two example network motifs. For these it only makes sense to perturb the non-zero entries $(k,l)$ such that $A_{k,l}\neq 0$, and only such that ${\rm sign}(\epsilon)={\rm sign}(A_{k,l})$.  For the trophic chain motif, $\Tri$, for each $k$ and $l$,
$$
\lim_{\epsilon\rightarrow \infty}{\rm \NS}\left(\Tri,{\rm sign}\left({\Tri}_{k,l}\right)\epsilon,k,l\right)=0.
$$
Hence, no sign switches will occur for any choice of non-zero $(k,l)$ entries; the trophic chain is entirely sign insensitive to error in its $A_{k,l}$ magnitudes (see also Fig. \ref{fig:QualSens}a). The reason for this is given in Theorem \ref{thm:TridiagonalInverse}.

In contrast, for the intraguild predation motif $\IGP$ the average (over all non-zero $k$ and $l$) fraction of sign switches is given by using Corollary \ref{cor:limitingbehavior} which results in:
$$
\lim_{\epsilon\rightarrow\infty}\frac{1}{12}\sum_{k,l\ {\rm s.t.}\ \IGP_{k,l}\neq 0}\frac{1}{4^2}{\rm \NS}\left(\IGP,{\rm sign}\left({\IGP}_{k,l}\right)\epsilon ,k,l\right)= 0.427,
$$
showing that for this matrix, perturbing a single entry leads to a higher average fraction of sign switches than is expected on average when using independent standard normal distributions for the entries. An explanation for this is given in Section \ref{section:IGPExample}.

\end{remark}

\subsection{How perturbation affects the norm of the inverse}
\label{section:Norm}
We now aim to investigate how perturbing a single entry in $A$ affects the norm of the inverse. {We do this for a few different norms, each useful in describing how \textit{quantitatively} different the inverse of the perturbed matrix is from the inverse of the unperturbed matrix (i.e. how quantitatively sensitive predictions of species responses are to error in $A$)}. Recall our standing assumptions that $A$ is invertible, for each $i$ and $j$, $A^{-1}_{i,j}\neq 0$, and when we consider a perturbation by another matrix $B$, we assume that $A+B$ is invertible.

\subsubsection{Entry-wise 1-norm}
We first investigate the effect perturbation has on the entry-wise (vectorized) 1-norm. We use the atypical notation $||A||_{\rm T}$ to emphasize that this means the \textit{total} absolute value of the entries of the matrix $A$, and to differentiate it from the operator 1-norm $||A||_1$.
\begin{definition}
For an $n\times m$ matrix $A$, let $$||A||_{\rm T}=\sum_{i=1}^n \sum_{j=1}^m |A_{i,j}|.$$
\end{definition}
A few other matrix norms will be helpful:
\begin{definition}
\label{def:spectralnorm}
Let $A$ be an $n \times m$ matrix, and $1\leq p \leq \infty$. Then the operator $p$-norm of $A$ is given by
$$||A||_p = \max_{x\neq 0} \frac{||Ax||_p}{||x||_p}$$
where $||\cdot||_p$ is the standard vector $p$-norm.
\end{definition}
Recall that $||A||_1=\max_{1\leq j\leq n} \sum_{i=1}^n |A_{i,j}|$ is the maximum absolute column sum of $A$ and $||A||_\infty=\max_{1\leq i\leq n} \sum_{j=1}^n |A_{i,j}|$ is the maximum absolute row sum of $A$.

As $|\epsilon|$ approaches infinity, the entry-wise 1-norm (total magnitude) of the difference between $A^{-1}$ and the perturbed matrix {(i.e. the difference in the total summed responses of all species)} reaches a (finite) fixed value. Indeed, using equation \eqref{eqn:perturbsingleentry} we find that
\begin{align}
||A^{-1}-(A+\epsilon \delta_{k,l})^{-1}||_{\rm T} & = \sum_{i,j=1}^n \left|A^{-1}_{i,j} - (A+\epsilon \delta_{k,l})^{-1}_{i,j}\right|\\
& = \sum_{i,j=1}^n \left| \frac{\epsilon A^{-1}_{i,k} A^{-1}_{l,j}}{1+\epsilon A^{-1}_{l,k}}\right|\\
& = \left|\frac{\epsilon}{1+\epsilon A^{-1}_{l,k}}\right| \sum_{i=1}^n |A^{-1}_{i,k}| \sum_{j=1}^n |A^{-1}_{l,j}|\\
&\xrightarrow[\epsilon \rightarrow \pm \infty]{}  \frac{1}{|A^{-1}_{l,k}|} \sum_{i=1}^n |A^{-1}_{i,k}| \sum_{j=1}^n |A^{-1}_{l,j}|\label{line:thisline}.
\end{align}

{This means that quantitative sensitivity is reduced by having the absolute value of the net effect of $k$ on $l$ be large relative to the sum of $k$'s absolute net effects on all species and the sum of the absolute net effects of all species on $l$.}

Further estimations are possible, including the observation that equation \eqref{line:thisline} implies that 
$$
\lim_{\epsilon \rightarrow \infty}||A^{-1}-(A+\epsilon \delta_{k,l})^{-1}||_{\rm T} \leq \frac{1}{|A^{-1}_{l,k}|} ||A^{-1}||_1 ||A^{-1}||_\infty.
$$
Similarly, if all entries of $A$ are nonzero, summing the relative total magnitude differences gives:
$$
\lim_{\epsilon \rightarrow \infty} \sum_{k,l} \frac{||A^{-1}-(A+\epsilon \delta_{k,l})^{-1}||_{\rm T}}{||A^{-1}||_{\rm T}} = \left\lVert A^{-1}\right\rVert_{\rm T} \left\lVert 1/A^{-1}\right\rVert_{\rm T}
$$
where by $1/A^{-1}$ we mean the matrix whose $(i,j)$ entry is equal to $\frac{1}{A^{-1}_{i,j}}$. This leads to the following measure of quantitative sensitivity of a matrix:
\begin{definition}
\label{def:MRS}
For an invertible $n\times n$ matrix $A$, let $M$ be the number of non-zero entries in $A$: $M=\left| \{(i,j)\ {\rm s.t.}\ A_{i,j}\neq 0\} \right|$. By the \textit{magnitude response sensitivity}, we mean
\begin{align}
{\MRS (A) =\frac{1}{M \cdot \lVert A^{-1} \rVert_{\rm T}} \sum_{k,l\ {\rm s.t.}\ A_{k,l}\neq 0}\frac{1}{|A^{-1}_{l,k}|} \sum_{i=1}^n |A^{-1}_{i,k}| \sum_{j=1}^n |A^{-1}_{l,j}|}.
\end{align}
\end{definition}
The quantity $\MRS(A)$ gives the mean (averaged over all individually perturbed entries) relative total magnitude difference between the inverse of the unperturbed matrix and the inverse of the matrix resulting from letting one entry be perturbed in an arbitrarily large fashion. Note that for matrices with no zero entries, the quantity $\MRS(A)$ is minimized for matrices $A$ with all entries equal in absolute value: any minimizer of the quantity $||A^{-1}||_{\rm T} ||1/A^{-1}||_{\rm T}$ has the property that $A^{-1}=a D$ for $a\in \mathbb{R}$ and $D$ an invertible matrix with entries in the set $\{-1,1\}$. Conversely, matrices $A$ with large variation (in absolute value) in their entries will exhibit large values of $\MRS(A)$, and hence larger quantitative sensitivity.

Using this definition, we can calculate the average (over all perturbed $k$ and $l$) total relative error between $A^{-1}$ and $\left(A+\epsilon \delta_{k,l}\right)^{-1}$. For $\Tri$, we have that
$$
\MRS\left(\Tri\right) = 3.53.
$$
This indicates that, on average, perturbing (to infinity) a single nonzero entry of $\Tri$ will result in the values of the inverse being off by factor of approximately $3.53$ when compared to the unperturbed matrix.
For $\IGP$, we find that
$$
\MRS\left(\IGP\right) = 1.63.
$$
This indicates that, on average, perturbing (to infinity) a single nonzero entry of $\IGP$ will result in the values of the inverse being off by a factor of only approximately $1.63$ when compared to the unperturbed matrix.
\begin{figure}[!h]%
\begin{center}
\includegraphics[width=3.25in,trim={0 0 0 0in},clip]{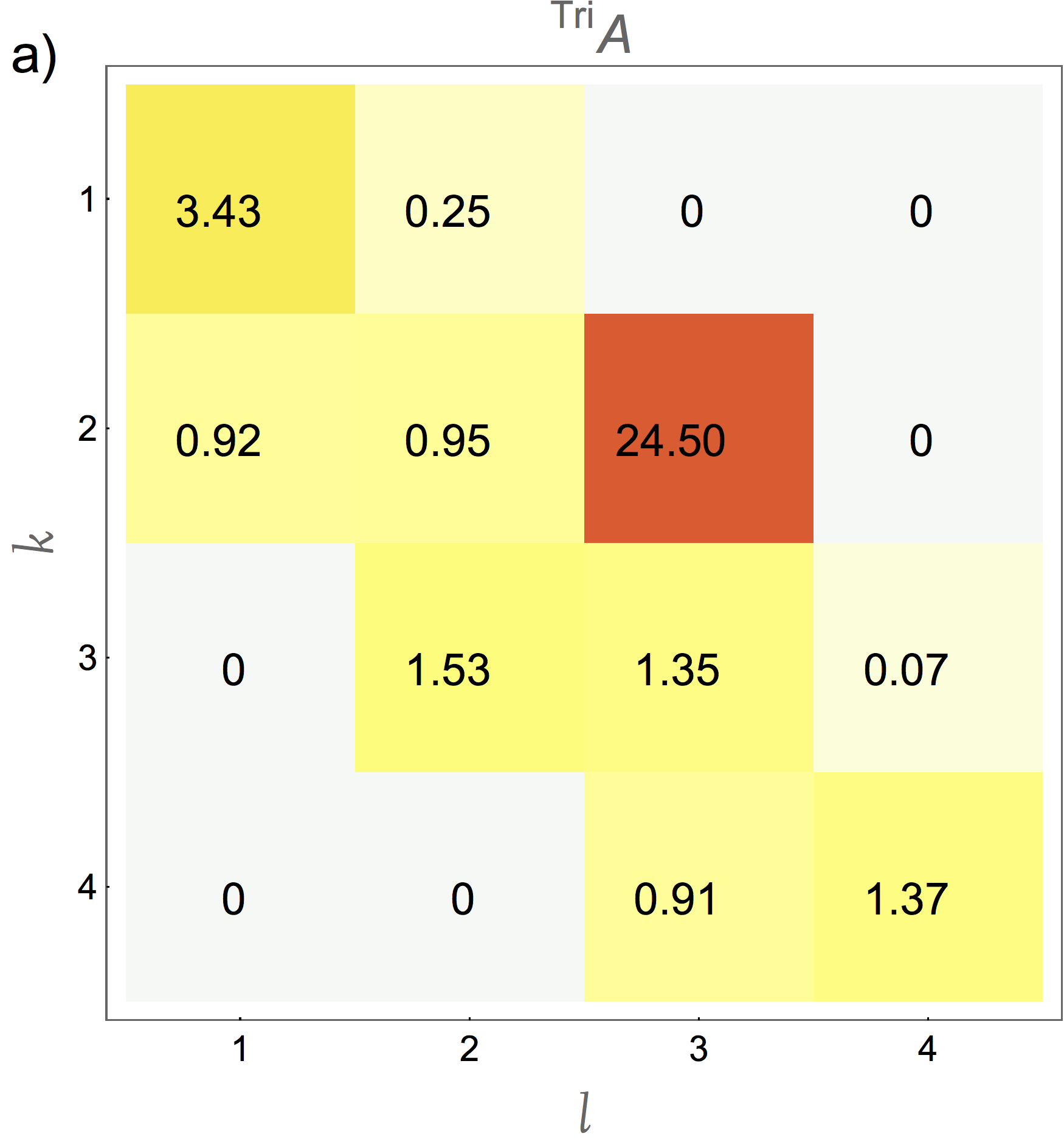}%
\includegraphics[width=3.25in,trim={0 0 0 0in},clip]{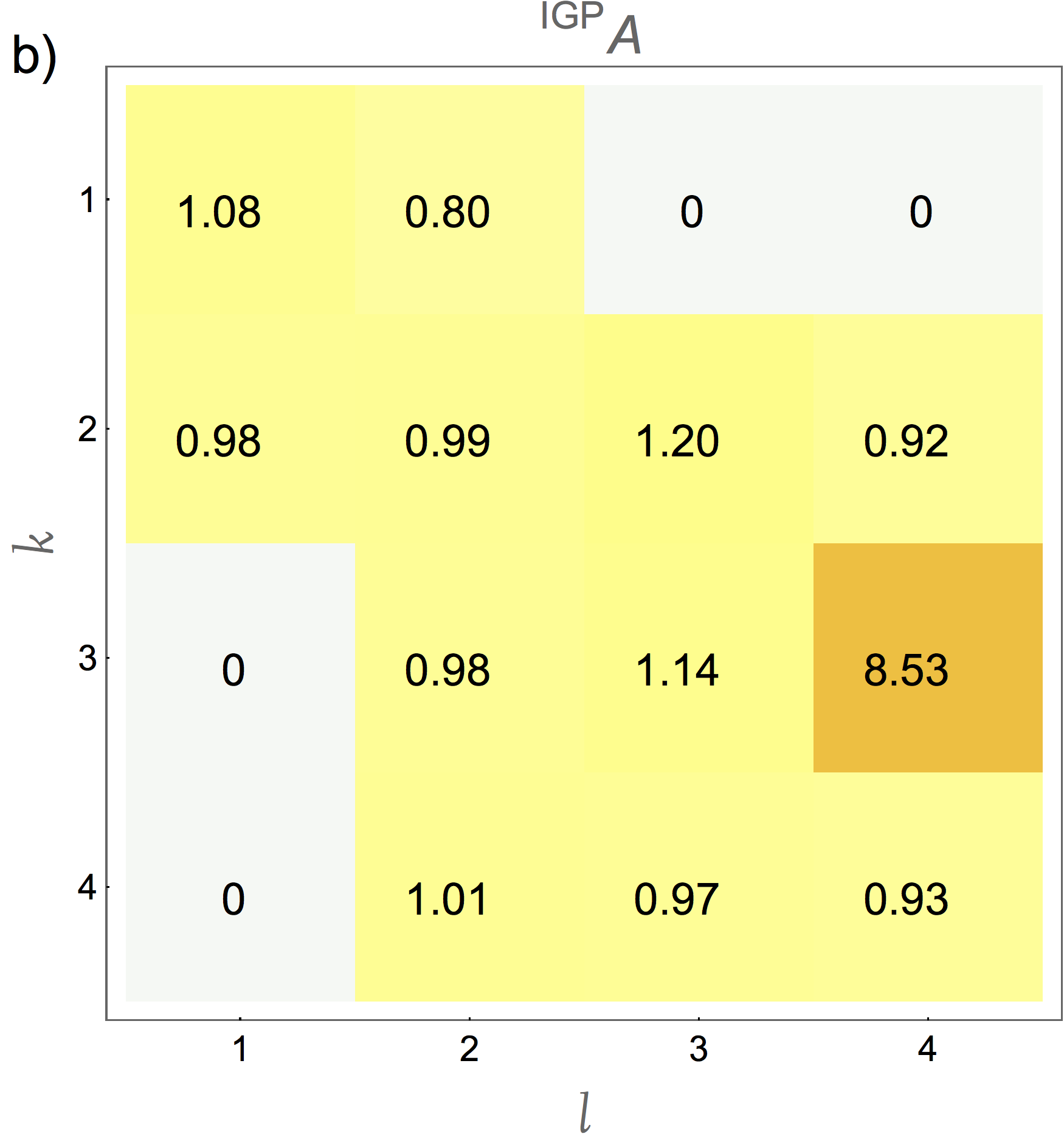}
\end{center}
\caption{Heat maps for the (quantitative) magnitude response sensitvity between the perturbed and unperturbed matrices: $\lim_{\epsilon \rightarrow \infty} \frac{||A^{-1}-(A+\epsilon \delta_{k,l})^{-1}||_{\rm T}}{||A^{-1}||_{\rm T}}$. a) The trophic chain motif. b) The intraguild predation motif. The average quantitative relative error of the intraguild predation motif is roughly half as large as the average quantitative relative error of the trophic chain. Compare this to their qualitative sensitivities shown in Figure \ref{fig:QualSens}.}
\label{fig:QuantSens}%
\end{figure}

Hence, while $\IGP$ is more sign sensitive to large $\epsilon$ perturbations (see the remark at the end of Section \ref{section:LimitingValue}), the matrix $\Tri$ exhibits significantly less total change in magnitude when perturbed; the trophic chain is more quantitatively sensitive to error than is the intraguild predation motif. Figure \ref{fig:QuantSens}, where the individual relative errors $\lim_{\epsilon \rightarrow \infty} \frac{||A^{-1}-(A+\epsilon \delta_{k,l})^{-1}||_{\rm T}}{||A^{-1}||_{\rm T}}$ are depicted in a heat map for each $k$ and $l$, visualizes which entries are the cause of this quantitative sensitivity.  For example, the quantitative dynamics of the trophic chain are clearly most sensitive to uncertainty in the top-down effect of species 3 (the intermediate consumer) on the primary consumer (species 2) (i.e. $A_{2,3}$), and relatively insensitive to uncertainty in the reciprocal interaction of these two species (i.e. $A_{3,2}$).

\subsubsection{Spectral Norm}
We now study how perturbing the matrix $A$ affects the spectral norm ($||~\cdot~||_2$ from Definition \ref{def:spectralnorm}) of the inverse matrix. {While the spectral norm has no clear ecological interpretation, our} goal is to relate the spectral norm of $A^{-1}-\left(A+\epsilon \delta_{k,l}\right)^{-1}$ with the singular values of $A$ {whose importance and properties are well known.} We first define our notation. 
\begin{definition}[Singular Values]
For a real matrix $A\in \mathbb{R}^{n \times m}$ the singular values $\sigma_i(A)$ are the square roots of the eigenvalues of $A^{T}A$ listed (with their multiplicities) in nonincreasing order $$\sigma_1(A) \geq \sigma_2(A) \geq \cdots \geq \sigma_n(A).$$
\end{definition}
We will at times write $\sigma_{\rm max}(A)$ and $\sigma_{\rm min}(A)$ for $\sigma_1(A)$ and $\sigma_n(A)$ respectively.
Recall that the spectral norm $||A||_2$ is equal to the dominant singular value: $||A||_2 = \sigma_1(A)$.
We will also have need of the Frobenius norm.
\begin{definition}[Frobenius norm]
For a real matrix $A\in \mathbb{R}^{m\times n}$, let $$||A||_F=\left(\sum_{i=1}^m \sum_{j=1}^n |A_{i,j}|^2\right)^{1/2}.$$
\end{definition}
\noindent
Recall that for any matrix $A\in \mathbb{R}^{m\times n}$, $||A||_F=\left(\sum_{i=1}^{\min\{m,n\}} \sigma_i(A)^2\right)^{1/2}$ and so $||A||_2\leq ||A||_F$ with equality if and only if $A$ is rank 1. 

We aim to sum the norm $\left|\left| A^{-1}-(A+\epsilon \delta_{k,l})^{-1}\right|\right|_2$ over all $k$ and $l$ entries to thereby estimate the Euclidean distances between the unperturbed and perturbed inverse matrices. We consider the ecologically realistic case where we only perturb the non-zero entries of $A$ and ensure that the perturbation to the $(k,l)$ entry does not change the sign of $A_{k,l}$. Hence, when perturbing the $(k,l)$ entry, let $t>0$ be a variable real number and define $\epsilon_{k,l}(t)={\rm sign}(A_{k,l}) t$, recalling that for a scalar $x$, ${\rm sign}(x)$ is equal to 1 if $x>0$, -1 if $x<0$, and 0 if $x=0$.
\begin{theorem}
\label{thm:UpperBound}
Let $A\in \mathbb{R}^{n\times n}$ be an invertible real matrix such that for each $k,l=1,\dots,n$, $A^{-1}_{k,l}\neq 0$. Let $t\in \mathbb{R}_{>0}$ and for each $k,l=1,\dots,n$, let $\epsilon_{k,l}(t) = {\rm sign}(A_{k,l}) t$. Then
$$
\lim_{t\rightarrow \infty} \sum_{k,l=1}^n \left|\left| A^{-1}-\left(A+\epsilon_{k,l}(t) \delta_{k,l}\right)^{-1}\right|\right|^2_2 \leq \max_{k,l} \frac{1}{|A^{-1}_{k,l}|^2} \left(\sum_{i=1}^n \frac{1}{\sigma_i(A)^2}\right)^2
$$
\end{theorem}
{This means that the maximum difference between the inverse of the perturbed and the inverse of the unperturbed matrix (as quantified by the spectral norm) is inversely related to the absolute values of the inverse of the unperturbed matrix and to the singular values of the unperturbed matrix.}
\begin{proof}
Observe that the matrix with $(i,j)$ entry equal to $\frac{\epsilon A^{-1}_{i,k} A^{-1}_{l,j}}{1+\epsilon A^{-1}_{l,k}}$, has rank equal to 1 as it is equal to a scalar times an outer product involving a column and a row of $A^{-1}$. Combining this fact with Theorem \ref{thm:Rank1Pert} allows us to relate the spectral norm of $A^{-1}-\left(A+\epsilon \delta_{k,l}\right)^{-1}$ with the Frobenius norm of the outer product of the $k^{\rm th}$ column and $l^{\rm th}$ row of $A^{-1}$. Calculating:
\begin{align}
\label{eqn:SpectralNormStart}
\sum_{k,l=1}^n \left|\left| A^{-1}-\left(A+\epsilon_{k,l}(t) \delta_{k,l}\right)^{-1}\right|\right|^2_2 &= \sum_{k,l=1}^n \left| \left| A^{-1} - \left( A^{-1} - \frac{\epsilon_{k,l}(t) A^{-1}_{i,k} A^{-1}_{l,j}}{1+\epsilon_{k,l}(t) A^{-1}_{l,k}}\right)\right|\right|_2^2\\
&=\sum_{k,l=1}^n \sum_{i,j=1}^n \left| \frac{\epsilon_{k,l}(t) A^{-1}_{i,k} A^{-1}_{l,j}}{1+\epsilon_{k,l}(t) A^{-1}_{l,k}} \right|^2 \\
& = \sum_{k,l=1}^n \left| \frac{\epsilon_{k,l}(t)}{1+\epsilon_{k,l}(t) A^{-1}_{l,k}} \right|^2 \sum_{i,j=1}^n \left| A^{-1}_{i,k} A^{-1}_{l,j} \right|^2\\
&\leq \max_{k,l} \left| \frac{\epsilon_{k,l}(t)}{1+\epsilon_{k,l}(t) A^{-1}_{l,k}} \right|^2 \sum_{k,l=1}^n \sum_{i,j=1}^n \left| A^{-1}_{i,k} A^{-1}_{l,j} \right|^2\\
&= \max_{k,l} \left| \frac{\epsilon_{k,l}(t)}{1+\epsilon_{k,l}(t) A^{-1}_{l,k}} \right|^2 \sum_{i,k=1}^n \left| A^{-1}_{i,k}\right|^2 \sum_{l,j=1}^n \left| A^{-1}_{l,j}\right|^2\\
&= \max_{k,l} \left| \frac{\epsilon_{k,l}(t)}{1+\epsilon_{k,l}(t) A^{-1}_{l,k}}\right|^2 \left|\left|A^{-1}\right|\right|^4_F\\
&= \max_{k,l} \left| \frac{\epsilon_{k,l}(t)}{1+\epsilon_{k,l}(t) A^{-1}_{l,k}}\right|^2 \left(\sum_{i=1}^n \sigma_i(A^{-1})^2\right)^2. \label{eqn:SpectralNormEnd}
\end{align}
Recalling that we defined  $\epsilon_{k,l}(t)={\rm sign}(A_{k,l})t$, we have
$$
\lim_{t\rightarrow \infty} \max_{k,l} \left| \frac{\epsilon_{k,l}(t)}{1+\epsilon_{k,l}(t) A^{-1}_{l,k}}\right|^2 = \lim_{t\rightarrow \infty} \max_{k,l} \left| \frac{t}{1\pm t A^{-1}_{l,k}}\right|^2 = \max_{k,l} \frac{1}{|A_{k,l}^{-1}|^2}.
$$
Finally, using the fact that $\sigma_i(A^{-1}) =1/\sigma_i(A)$ for all $i$, and taking the limit as $t\rightarrow \infty$, the result follows.
\end{proof}
In the case where no entry of $A$ is zero, we have a lower bound as well.
\begin{theorem}
\label{thm:LowerBound}
Let $A\in \mathbb{R}^{n\times n}$ be an invertible real matrix such that for each $k,l=1,\dots,n$, both $A_{k,l}\neq 0$ and $A^{-1}_{k,l}\neq 0$. Let $t\in \mathbb{R}_{>0}$ and for each $k,l=1,\dots,n$, let $\epsilon_{k,l}(t) = {\rm sign}(A_{k,l}) t$. Then
$$
\lim_{t\rightarrow \infty} \sum_{k,l=1}^n \left|\left| A^{-1}-\left(A+\epsilon_{k,l}(t) \delta_{k,l}\right)^{-1}\right|\right|^2_2 \geq \min_{k,l} \frac{1}{|A^{-1}_{k,l}|^2} \frac{1}{\sigma_{\rm min}(A)^4}.
$$
\end{theorem}
\noindent
Therefore, the minimum difference between the inverse of the perturbed and the inverse of the unperturbed matrix (as quantified by the spectral norm) is also inversely related to the absolute values of the inverse of the unperturbed matrix (as in Theorem \ref{thm:UpperBound}) and to the smallest singular value of the unperturbed matrix.
\begin{proof}
The proof follows from a calculation similar to that in the proof of Theorem \ref{thm:UpperBound}:
\begin{align}
\sum_{k,l=1}^n \left| \left| A^{-1}-\left(A+\epsilon_{k,l}(t) \delta_{k,l}\right)^2 \right| \right|_2^2 &= \sum_{k,l=1}^n \sum_{i,j=1}^n \left| \frac{\epsilon_{k,l}(t) A^{-1}_{i,k} A^{-1}_{l,j}}{1+\epsilon_{k,l}(t) A^{-1}_{l,k}}\right|^2\\
&= \sum_{k,l=1}^n \left| \frac{\epsilon_{k,l}(t)}{1+\epsilon_{k,l}(t) A^{-1}_{l,k}}\right|^2 \sum_{i,j=1}^n |A^{-1}_{i,k}|^2 |A^{-1}_{l,j}|^2\\
&\geq \min_{k,l} \left| \frac{\epsilon_{k,l}(t)}{1+\epsilon_{k,l}(t) A^{-1}_{l,k}}\right|^2 ||A^{-1}||^4_F\\
&\geq \min_{k,l} \left| \frac{\epsilon_{k,l}(t)}{1+\epsilon_{k,l}(t) A^{-1}_{l,k}}\right|^2 ||A^{-1}||^4_2\\
&=\min_{k,l} \left| \frac{\epsilon_{k,l}(t)}{1+\epsilon_{k,l}(t) A^{-1}_{l,k}}\right|^2 \sigma_{\rm max}(A^{-1})^4\\
&=\min_{k,l} \left| \frac{\epsilon_{k,l}(t)}{1+\epsilon_{k,l}(t) A^{-1}_{l,k}}\right|^2 \frac{1}{\sigma_{\rm min}(A)^4}
\end{align}
Since each entry of $A$ is nonzero, this implies that for all $k,l$, $\epsilon_{k,l}(t) = {\rm sign}(A_{k,l})t$, and hence $\lim_{t\rightarrow \infty} \min_{k,l} \left| \frac{\epsilon_{k,l}(t)}{1+\epsilon_{k,l}(t) A^{-1}_{l,k}}\right|^2>0$. Taking limits as $t\rightarrow \infty$ leads to the result.
\end{proof}
\subsection{Perturbing Multiple Entries}
\label{section:PerturbManyEntries}
Up to this point we have assumed that only a single entry of $A$ is perturbed at a time.  However, in most applications, multiple if not all entries will have some level of uncertainty associated. Fortunately, Theorem \ref{thm:Rank1Pert} can be applied iteratively to compute $(A+B)^{-1}$ when $B$ is of rank $r$ by writing $B$ as a sum of rank 1 matrices. This leads to the following result:

\begin{theorem}[Theorem 1 of\cite{miller1981inverse}]
\label{thm:RankRPert}
Let $A$ and $A+B$ be nonsingular matrices where $B=B_{(1)}+\cdots+B_{(r)}$ has rank $r$ and each $B_{(i)}$ has rank 1 for $i=1,\dots,r$. Let $C_{(k+1)}=A+B_{(1)}+\cdots+B_{(k)}$ for $k=1,\dots,r$ and $C_{(1)}=A$. Then with $g_{(k)}=\frac{1}{1+{\rm tr}(C_{(k)}^{-1} B_{(k)})}$,
$$
C_{(k+1)}^{-1} = C_{(k)}^{-1} - g_{(k)} C_{(k)}^{-1} B_{(k)} C_{(k)}^{-1}.
$$
In particular,
$$
(A+B)^{-1} = C_{(r)}^{-1} - g_{(r)} C_{(r)}^{-1}B_{(r)}C_{(r)}^{-1}.
$$
\end{theorem}

Utilizing Theorem \ref{thm:RankRPert} allows one to programmatically compute the inverse of $(A+B)^{-1}$, an arbitrary set of perturbations of the matrix $A$ with any entry $B_{i,j}=\epsilon_{i,j}$ being either a fixed real number or a suitably chosen random variable. This allows results similar to Lemma \ref{lemma:PerturbSingle} to be obtained through the assistance of a computer algebra system. Similarly, formulas for the generalization of the number of switches ${\rm \NS}$ and the expected number of switches $\mathbbm{E}({\rm \NS})$, while notationally unwieldy, are computationally straightforward to utilize, as they still only depend on the entries of $A^{-1}$ and the values of $B_{i,j}=\epsilon_{i,j}$.

\subsubsection{Index of Sign Sensitivity}
\label{section:IndexOfStability}
We next perturb more than one entry to obtain an index of sign sensitivity which characterizes the overall proclivity of an arbitrary perturbation to cause sign switches in the inverse of the perturbed matrix. We motivate this using our example motifs, recalling for convenience that the definition of $\IGP$ given in equation \eqref{IGPMatrix} is:
\begin{align}
\label{eqn:ExampleStabA}
\IGP=\left(
\begin{array}{cccc}
 -0.237 & -1 & 0 & 0 \\
 0.1 & -0.015 & -1 & -1 \\
 0 & 0.1 & -0.015 & -1 \\
 0 & 0.045 & 0.1 & -0.015 \\
\end{array}
\right).
\end{align}

We first consider a perturbation to $\IGP$ of $B$, defined by
\begin{align}
B=\left(
\begin{array}{cccc}
 0 & 0 & 0 & 0 \\
 0 & 0 & 0 & 0 \\
 0 & 0 & 0 & 0 \\
 0 & \epsilon_{4,2} & \epsilon_{4,3} & 0 \\
\end{array}
\right),
\end{align}
where $\epsilon_{4,2}$ ranges uniformly over the interval $[-0.045, 0.955]$ and $\epsilon_{4,3}$ ranges uniformly over the interval $[-.1, .9]$ so as not to reverse their sign. Even though this $B$ is still a rank 1 perturbation, it is ecologically relevant and is illustrative to consider. This represents the case where all entries of $\IGP$ are known exactly, excepting the $(4,2)$ and $(4,3)$ entries, and where the signs of these entries are known, but where their magnitudes are allowed to vary over an interval of arbitrary length 1. Applying Theorem \ref{thm:RankRPert} allows one to obtain a (large) formula for the inverse of $\IGP+B$.

\begin{figure}[!h]%
\begin{center}
\includegraphics[width=3.25in,trim={0 0 0 0in},clip]{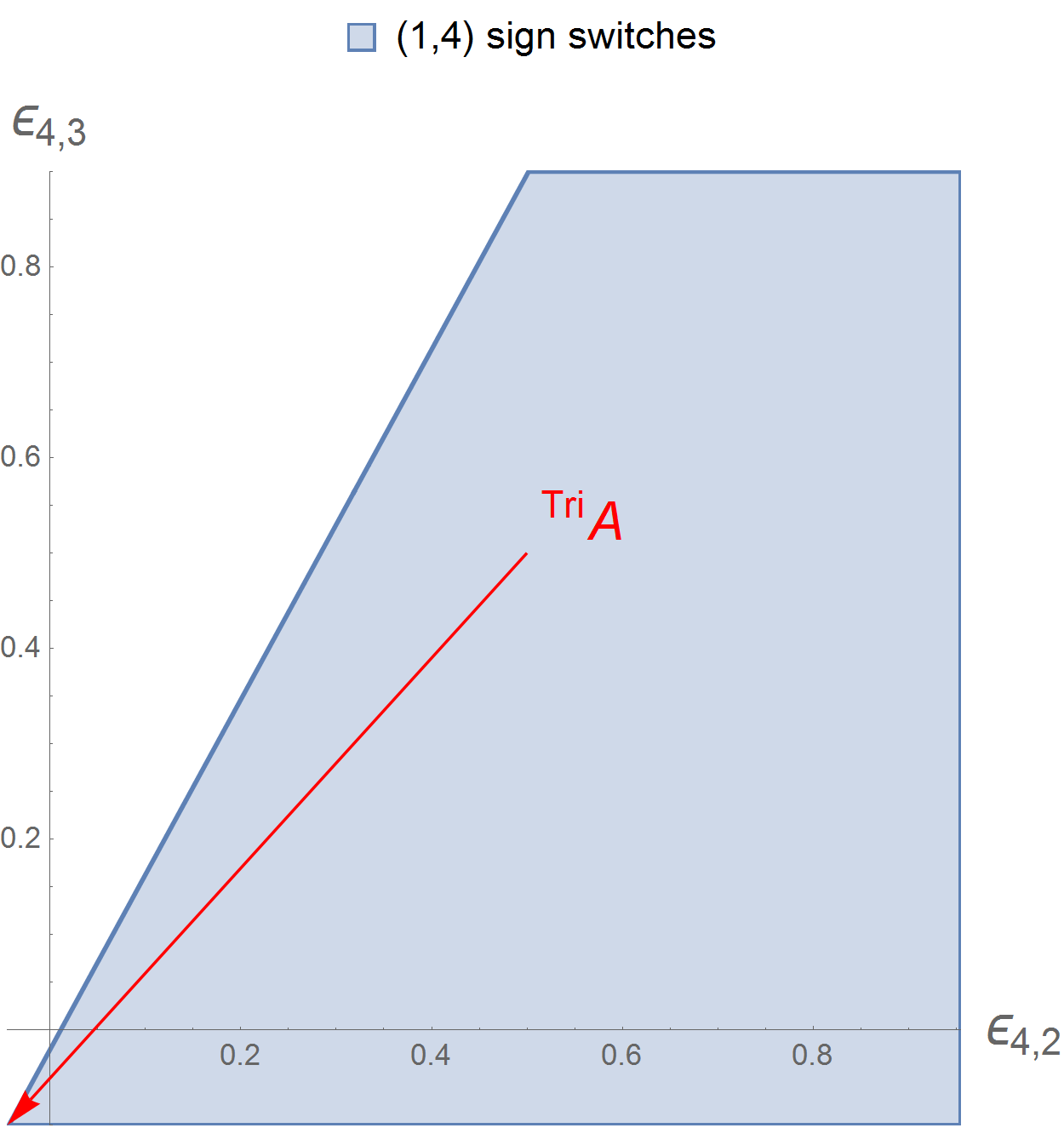}%
\end{center}
\caption{Depiction of the region where the perturbation values $\epsilon_{4,2}$ and $\epsilon_{4,3}$ cause a sign switch in the $(1,4)$ entry of $\left(\IGP+B\right)^{-1}$ in comparison to the $(1,4)$ entry of $\left(\IGP\right)^{-1}$. {Note that trophic chain motif $\Tri$ corresponds to the highlighted special case where the two errors are of equal magnitude but opposite sign to their respective entries in $\IGP$. }}
\label{fig:TwoRegionOfStab}%
\end{figure}

To illustrate, we focus on the $(1,4)$ entry of $\left(\IGP+B\right)^{-1}$ and compare this to the $(1,4)$ entry of $\left(\IGP\right)^{-1}$ to see what values of $\epsilon_{4,2}$ and $\epsilon_{4,3}$ cause the sign of this entry to switch. Utilizing a computer algebra system, this reduces to the following inequality: 
\begin{align}
\label{eqn:ineqs}
\epsilon_{4,3} < -0.0204257 + 1.83447 \epsilon_{4,2}.
\end{align}
Plotting the region where this expression is true leads to a depiction of the \textit{region of sign sensitivity} (i.e. sign indeterminacy): the region where values of $\epsilon_{4,2}$ and $\epsilon_{4,3}$ cause a sign switch in the $(1,4)$ entry of the perturbation $\left(\IGP+B\right)^{-1}$ when compared to the $(1,4)$ entry of $\left(\IGP\right)^{-1}$ (Figure \ref{fig:TwoRegionOfStab}).

For general $A$, we can generalize the approach to where $B_{ij}=\epsilon_{ij}$ is a matrix of same size as $A$ with each $\epsilon_{i,j}$ taking on values in a given interval (while still preserving the sign of $A+B$ in comparison to $A$) and consider the resulting inequalities for all entries of $(A+B)^{-1}_{ij}$. Comparing the volume of the region where this systems of inequalities is not all true to the total volume of the perturbation space leads to an index of \textit{sign sensitivity}:
\begin{definition}
\label{def:SwitchStability}
Given a fixed invertible $m\times n$ matrix $A$ and an $m\times n$ matrix with entries $B_{i,j}=\epsilon_{i,j}$ whose values range over the region $\mathscr{R}\subset \mathbbm{R}^{m+n}$, let $\mathscr{S}$ be the subregion in $\mathscr{R}$ satisfying:
\begin{align}
(\epsilon_{1,1},\cdots,\epsilon_{m,n})\in \mathscr{S} \iff \exists i,j, (A+B)^{-1}_{i,j} \text{ has the opposite sign as } A^{-1}_{i,j}.
\end{align}
Then define the sign sensitivity of $A$ as
\begin{align}
\label{eqn:SwitchStability}
{\rm \SSS}(A) = \frac{{\rm vol}(\mathscr{S})}{{\rm vol}(\mathscr{R})}
\end{align}
\end{definition}
\noindent
This index is interpreted as the percentage of perturbation space in which some entry of the perturbed matrix inverse $(A+B)^{-1}$ changes sign in comparison to $A^{-1}$.  This definition can be extended by incorporating a probability distribution over $\mathscr{R}$ (and subsequently $\mathscr{S}$ as well):
\begin{definition}
\label{def:SwitchStabilityProb}
Given a fixed invertible $m\times n$ matrix $A$ and an $m\times n$ matrix with entries $B_{i,j}=\epsilon_{i,j}$ whose values range over the region $\mathscr{R}\subset \mathbbm{R}^{m+n}$ according to the multivariate distribution $\mathscr{D}$ supported on $\mathscr{R}$. 
Let $\mathscr{S}$ be the subregion in $\mathscr{R}$ satisfying:
\begin{align}
(\epsilon_{1,1},\cdots,\epsilon_{m,n})\in \mathscr{S} \iff \exists i,j, (A+B)^{-1}_{i,j} \text{ has the opposite sign as } A^{-1}_{i,j}.
\end{align}
Then define the distributional sign sensitivity of $A$ as
\begin{align}
\label{eqn:SwitchStabilityProb}
{\rm \SSS}_\mathscr{D}(A) = \mathbbm{P}_{\mathscr{D}}(\mathscr{S})
\end{align}
\end{definition}
Due to Theorem \ref{thm:RankRPert} and the subsequent discussion, both ${\rm \SSS}(A)$ and ${\rm \SSS}_\mathscr{D}(A)$ can be computed explicitly and hence efficiently via Monte-Carlo sampling.

For example, for $\mathscr{D}$ being the product of independent uniform distributions over intervals of length 1 that do not change the signs of the original matrix $A$, we calculate the distributional sign sensitivity of $\IGP$ to be 
$$
{\rm \SSS}_{\mathscr{D}}\left(\IGP\right)\approx 0.947,
$$
indicating that approximately $94.7\%$ of all perturbation values incur sign changes in $\left(\IGP\right)^{-1}$ due to this error distribution. Equivalently, this means if each $\epsilon_{i,j}$ is a uniform random variable in the interval $[-\IGP_{i,j}, 1-\IGP_{i,j}]$ for $\IGP_{i,j}>0$ and $[-1-\IGP_{i,j},-\IGP_{i,j}]$ for $\IGP_{i,j}<0$, and a point mass on 0 if $\IGP_{i,j}=0$, then the probability of at least one sign switch occurring in $\left(\IGP+B\right)^{-1}$ in comparison to $\left(\IGP\right)^{-1}$ is equal to $0.947$.

In comparison, for the trophic chain motif $\Tri$ and {a similarly} defined distribution $\mathscr{D}$, 
$$
{\rm \SSS}_{\mathscr{D}}\left(\Tri\right)=0.
$$
This corresponds to Theorem \ref{thm:TridiagonalInverse} wherein we show that for this topology, the signs of the inverse do not depend on the magnitudes of the $A_{i,j}$ entries. Note that, just as in Corollary \ref{cor:Gauss}, the index of distributional sign sensitivity may be applied using any other distributions of errors as well.

\section{The Sign Sensitivity of Tridiagonal Matrices}
\label{section:TriDiag}
The trophic chain motif corresponds to a tridiagonal matrix.  We will subsequently utilize this fact for decomposing networks to understand their sensitivity.  In this Section we therefore investigate the qualitative sensitivity of tridiagonal matrices.

A number of authors have derived explicit formulas for the inverse of a tridiagonal matrix \cite{usmani1994inversiona,usmani1994inversionb,Fonseca2007eigenvalues,lewis1982inversion}. We use the notation of \cite{Fonseca2007eigenvalues} for the following Theorem:
\begin{theorem}[Lemma 3 of \cite{usmani1994inversionb}]
\label{thm:Tinverse}
Given an $n\times n$ nonsingular tridiagonal matrix 
\begin{align}
T = \left(\begin{array}{ccccc}
a_1 & b_1 & & &\\
c_1 & a_2 & b_2 & &\\
    & c_2 & \ddots & \ddots & \\
	&  & \ddots    & \ddots &  b_{n-1}\\
		&     &        & c_{n-1} & a_n
\end{array}
\right)
\end{align}
Let $\theta_i$ satisfy the recurrence relation
\begin{align}
\theta_i = a_i \theta_{i-1} - b_{i-1} c_{i-1} \theta_{i-2},\quad i=2,\dots,n
\end{align}
with initial conditions $\theta_0=1$, $\theta_1=a_1$.
Let $\psi_i$ satisfy the recurrence relation
\begin{align}
\psi_i = a_i \psi_{i+1} - b_i c_i \psi_{i+2}, \quad i=n-1,\dots,1
\end{align}
with initial conditions $\psi_{n+1} = 1$, $\psi_n = a_n$. Then
\begin{align}
\left(T^{-1}\right)_{i,j} = \left\{\begin{array}{ll}
(-1)^{i+j} b_i \cdots b_{j-1} \theta_{i-1} \psi_{j+1}/\theta_n & {\rm if}\ i\leq j\\
(-1)^{i+j} c_j \cdots c_{i-1} \theta_{j-1} \psi_{i+1}/\theta_n & {\rm if}\ i> j.
\end{array}
\right.
\end{align}
\end{theorem}

Using this result, we can demonstrate when no entry of $T^{-1}$ will change sign as the elements of $T$ are perturbed. 
For a given matrix $A$, by $\sign(A)$ we mean a matrix with $(i,j)^{\rm th}$ entry equal to $\sign(A_{i,j})$.
We now show that for certain tridiagonal matrices $T$, the sign pattern of $T^{-1}$ depends only on the sign pattern of $T$ and not on the magnitudes of the entries of $T$.
\begin{theorem}
\label{thm:TridiagonalInverse}
Let $T$ be a nonsingular tridiagonal matrix as in Theorem \ref{thm:Tinverse} such that for all $i$ and $j$, if $a_i\neq 0$ and $a_j\neq 0$, then ${\rm sign}(a_i)={\rm sign}(a_j)$, and if $c_i\neq 0$ and $b_i\neq 0$,  then ${\rm sign}(c_i)=-{\rm sign}(b_i)$.
Then $\sign(T^{-1}_{i,j})$ can be calculated directly from $\sign(T)$. 
\end{theorem}
\noindent
In other words, the qualitative net effects between species in a trophic chain motif are determined completely by the topology of the motif and not by the magnitudes of the interaction strengths. 

\begin{proof}
We sketch the proof for the case where for all $i$, $a_i\leq 0$, $b_i\leq 0$, and $c_i\geq 0$ as the other cases proceed similarly.
Using the same notation as in equation \eqref{thm:Tinverse},
\begin{align}
\theta_1 &= a_1 \leq 0\\
\theta_2 &= a_2 \theta_1 - c_1 b_1 \theta_0 = \overbrace{a_2 a_1}^\text{positive} - \overbrace{c_1 b_1}^\text{negative} \geq 0\\
\theta_3 &= \overbrace{a_3 \theta_2}^\text{negative} -\ \overbrace{c_2 b_2 \theta_1}^\text{positive} \leq 0.
\end{align}
Continuing in this fashion, one can see that 
\begin{align}
\sign(\theta_i) = \left\{ \begin{array}{ll} 1 & {\rm if}\ i\ {\rm is\ even}\\ -1 & {\rm if}\ i\ {\rm is\ odd}\end{array}\right..
\end{align}
Proceeding in a similar fashion for $\psi_i$, one can observe that
\begin{align}
\sign(\psi_i) = \left\{ \begin{array}{ll} 1 & {\rm if}\ i\not \equiv n \mod 2 \\ -1 & {\rm if}\ i\equiv n \mod 2\end{array}\right..
\end{align}
Thus, since $T^{-1}_{i,j}$ is equal to a product of terms whose signs do not depend on the magnitude of $a_k$, $b_k$, or $c_k$, the Theorem immediately follows.
\end{proof}

For the matrix $\Tri$, this implies that if we perturb the entries of $\Tri$ without changing their signs, the resulting inverse will have the same sign pattern as that of the inverse of the unperturbed matrix: ${\rm sign}\left(\left(\Tri\right)^{-1}\right)$. This indicates why we previously found that the sign sensitivity ${\rm \SSS}_{\mathscr{D}}\left(\Tri\right)=0$, and that for any $k,l$, the value $\lim_{\epsilon \rightarrow \infty}{\rm \NS}\left(\Tri,{\rm sign}\left(\Tri_{k,l}\right)\epsilon,k,l\right)=0$.   This insight extends to any linear chain of species interactions, including the motifs of apparent competition and exploitative competition \cite{Holt:1977kx}.

\section{The Sign Sensitivity of the IGP motif}
\label{section:IGPExample}
{In Section \ref{section:PerturbManyEntries} it was seen that even slight perturbations of the entries of the $\IGP$ matrix caused sign switches in it inverse (i.e. the sign sensitivity ${\rm \SSS}_{\mathscr{D}}(\IGP)\approx 0.947$). We here provide the means to understand why this is so by combining the use of Theorems \ref{thm:Rank1Pert} and \ref{thm:Tinverse}.}

For convenience, recall that 
\begin{align}
\IGP=\left(
\begin{array}{cccc}
 -0.237 & -1 & 0 & 0 \\
 0.1 & -0.015 & -1 & -1 \\
 0 & 0.1 & -0.015 & -1 \\
 0 & 0.045 & 0.1 & -0.015 \\
\end{array}
\right).
\end{align}
This matrix can be written as the sum of tridiagonal matrix plus a rank one matrix:
\begin{align}
\IGP&=\left(
\begin{array}{cccc}
 -0.237 & -1 & 0 & 0 \\
 0.1 & 0.435 & -1 & 0 \\
 0 & 0.1 & -0.015 & -1 \\
 0 & 0 & 0.1 & -1.015 \\
\end{array}
\right)
+
\left(\begin{array}{c}
0\\
-1\\
0\\
1
\end{array}
\right)
\cdot
\left(\begin{array}{cccc}
0 & 0.045 & 0 & 1
\end{array}
\right)
.
\end{align}
Let $T$ be the tridiagonal matrix in this decomposition:
\begin{align}
T&=\left(
\begin{array}{cccc}
 -0.237 & -1 & 0 & 0 \\
 0.1 & 0.435 & -1 & 0 \\
 0 & 0.1 & -0.015 & -1 \\
 0 & 0 & 0.1 & -1.015 \\
\end{array}
\right) 
\end{align}
and let $B$ be the rank one matrix which is the outer product of $u$ and $v$,
\begin{align}
u&=\left(\begin{array}{c}
0\\
-1\\
0\\
1
\end{array}
\right)\\
v'&=\left(\begin{array}{cccc}
0 & 0.045 & 0 & 1
\end{array}
\right)\\
B&=uv'=
\left(
\begin{array}{cccc}
 0 & 0. & 0 & 0 \\
 0 & -0.045 & 0 & -1 \\
 0 & 0. & 0 & 0 \\
 0 & 0.045 & 0 & 1 \\
\end{array}
\right).
\end{align}
The matrix $B$ can therefore be thought of as a perturbation to a certain subset of entries in $A$, a situation that can be analyzed by the techniques developed in Section \ref{section:PerturbManyEntries}.  We therefore also have $\IGP = T + B$, such that the matrix $\IGP$ can be viewed as a perturbation of the tridiagonal matrix $T$. Observe that Theorem \ref{thm:Tinverse} indicates that the signs of the entries of $T^{-1}$ depend on the magnitude of the entries of $T$. This is due critically to the entry $T_{2,2}=0.435$ which causes the entries on the main diagonal of $T$ to \textit{not} all have the same sign. It is because $T_{2,2}$ is positive that $T$ is sign sensitive, as seen in the proof of Theorem \ref{thm:TridiagonalInverse}. The addition of the rank 1 perturbation $B$ that makes $\IGP$ even more sign sensitive (Section \ref{section:SignSwitches}). Indeed, the decomposition of $\IGP$ into $T$ and $B$ reveals that the sign sensitivity of $\IGP$ would be most reduced by having the magnitude of $T_{4,2}$ be less than the magnitude of $T_{2,2}$ as then only the sign sensitivity of $B$ would be of consequence.  In contrast, perturbations to the $T_{2,4}$ entry will not affect the sign sensitivity of $\IGP$ because $T_{2,4}$ has the same sign as $T_{4,4}$.  In ecological terms, the sign sensitivity of the intraguild predation motif is driven by the sensitivity of the bottom-up effect of the basal resource on the top consumer rather than by the reciprocal top-down effect, which is consistent with our  observations in Figure \ref{fig:QualSens}.

More generally, utilizing Theorem \ref{thm:Rank1Pert}, we have a formula for the inverse of $\IGP$:
\begin{align}
\left(\IGP\right)^{-1} = (T+B)^{-1} = T^{-1} - \frac{1}{1+v'T^{-1} u} T^{-1} v v' T^{-1}.
\end{align}
Since the signs of $T^{-1}$ depend on the magnitudes of the entries of $T$, the above formula indicates that the signs of $\left(\IGP\right)^{-1}$ must also depend on the magnitudes of the entries of $\IGP$.  This is why the signs of the inverse of $\IGP$ are sensitive to uncertainty in the elements of the original matrix.

\section{Conclusions}
The primary contributions of this work are four indices that characterize the qualitative and quantitative sensitivity of press perturbation responses to uncertainty (or intrinsic variation) in a system's interaction strengths: 
\begin{enumerate}
\item ${\rm \NS}(A,\epsilon,k,l)$, which denotes the number of sign switches incurred by an estimation error of magnitude $\epsilon$ to the ($k,l$) entry of $A$,
\item the magnitude response sensitivity $\rm MRS(A)$, which quantifies the relative total magnitude difference between a system's true press perturbation responses and those predicted with estimation error, and 
\item the sign sensitivity indices $\rm SS (A)$ and ${\rm \SSS}_{\mathscr{D}}(A)$, which quantify the percentage of possible error magnitude space in which at least one sign switch will occur, with error respectively considered as either a volume or a distribution of magnitudes.
\end{enumerate}
The ${\rm \NS}(A,\epsilon,k,l)$ index, which underlies much of our approach, enables one to identify the most sensitive interactions within the network which must be estimated most accurately to produce qualitatively robust predictions. Importantly, these indices are exact and may be computed with relative ease, thereby obviating the need for simulations in assessing the indeterminacy of complex networks.  A key advance is the separation of the estimation error magnitudes from their frequency distribution. Together with recent statistical advances in the empirical characterization of uncertainty in ecological networks \cite{Wolf:2015fk, Poisot:2016fk}, our study thereby bridges the way to probabilistic predictions of ecosystem dynamics \cite{Novak:2016zr}.

Applications of the indices to two well-studied food web motifs provide a proof-of-concept demonstration for how the indeterminacy of larger, truly complex networks may be decomposed and understood.  They also illustrate the useful insights the indices can provide. For example, the observation that the trophic chain realization is more quantitatively sensitive to uncertainty than is the intraguild predation realization, despite the trophic chain being entirely sign insensitive to any error, may be ascribed to the top-down link between the intermediate consumer and the secondary consumer (i.e. $\Tri_{2,3}$, Fig. \ref{fig:QuantSens}a). It is this interaction that will need to be estimated most accurately for accurate quantitative predictions to be made. Similarly, our approach also elucidates why the qualitative dynamics of the IGP realization are almost completely sign insensitive to uncertainty in the top-down direct effect of the top-predator on the shared resource ($\IGP_{2,4}$), but extremely sign sensitive to uncertainty in the reciprocal bottom-up effect of the shared resource on the top-predator ($\IGP_{4,2}$) as shown in Figure \ref{fig:QualSens}b), and explained in Section \ref{section:IGPExample}). This particular insight indicates that qualitative predictions for this motif are more sensitive to error in the characterization of the top-predator's numerical response (i.e. the efficiency with which the predator converts consumed prey to offspring) than to error in the characterization of its functional response (i.e. the rate at which the predator consumes prey).  Future work will need to address the degree to which this conclusion is dependent on the specific choice of parameters used in our example, or is general to the structure of the motif itself.

More generally, our theorems highlight how the analysis of qualitative models  (i.e. Loop Analysis)  represents a special case of the analysis of quantitative models using our approach.  As evidenced in Figure \ref{fig:TwoRegionOfStab}, the assessment of alternative network topologies reflects the case in which the assumed magnitudes of the $\epsilon_{i,j}$ errors correspond exactly to the negative magnitudes of their respective $A_{i,j}$ elements.  Likewise, the addition of new links corresponds to the perturbation of the zero entries of $A$.  More specifically, alternative network topologies reflect alternative perturbations of each other, as illustrated in Section \ref{section:IGPExample} by the decomposition of the intraguild predation motif into the trophic chain motif.  The ability to characterize the region of parameter space within which no sign switches occur by analytical means (see Section \ref{section:PerturbManyEntries}) will thus be particularly useful in applications where quantitative estimates of interaction strengths are unavailable, enabling a more robust determination of the consequences of characterizing interactions by only their sign and a unit-magnitude strength.  

It is important to note that our treatment of indeterminacy has not considered the stability and feasibility of the perturbed matrix.  In interpreting the sensitivities of the $\Tri$ and $\IGP$ motifs we have thereby ignored the requirement that no bifurcations are affected by the addition of errors via $B$, with all eigenvalues of $A+B$ remaining negative and the abundances of all but the press-perturbed species remaining positive \cite{Novak:2016zr}.  The network properties that control these important attributes of complex networks remain an active area of research \cite{Allesina:2012uq, Rohr:2014uq} but, for our approach, represent only additional sets of constraints to be imposed on the permissible distributions of the errors.  That these attributes are nonetheless directly related to a network's quantitative sensitivity is evidenced by the influence of the singular values in determining the upper and lower bounds of the spectral norm difference between the inverse of the perturbed and unperturbed matrices (Theorems \ref{thm:LowerBound} and \ref{thm:UpperBound}).  Indeed, this observation suggests that a network's quantitative sensitivity may be of empirical use as an early warning indicator of bifurcation events across which qualitative predictions will surely be inaccurate \cite{Novak:2016zr, Scheffer:2012fk}.

\clearpage
\bibliography{library}{}
\bibliographystyle{abbrv}

\end{document}